\documentclass[usenames,dvipsnames]{article}
\usepackage{arxiv}

\usepackage{lineno,hyperref}
\modulolinenumbers[5]

\usepackage[english]{babel} % Specify a different language here - english by default
\usepackage{url} % Required to insert dummy text. To be removed otherwise

\usepackage{mathptmx}
\usepackage{times}
\usepackage{textcomp}
\usepackage{gensymb} % degree. ie 360º
\usepackage{comment}
\usepackage{multirow}
\usepackage[framemethod=tikz]{mdframed}
\usepackage[labelfont=bf]{subcaption}

\usepackage{tcolorbox}% http://ctan.org/pkg/tcolorbox
\definecolor{mycolor}{rgb}{0,0,0}% Rule colour
\makeatletter
\newcommand{\mybox}[1]{%
	\setbox0=\hbox{#1}%
	\setlength{\@tempdima}{\dimexpr\wd0+13pt}%
	\begin{tcolorbox}[colframe=mycolor,boxrule=0.5pt,arc=4pt,
		left=6pt,right=6pt,top=6pt,bottom=6pt,boxsep=0pt,width=\@tempdima]
		#1
	\end{tcolorbox}
}

\usepackage{graphicx}

\usepackage{amsmath}
\usepackage{amsfonts}
\usepackage{amssymb}

\usepackage{xspace}

\usepackage{tcolorbox}

\usepackage{pgfpages,tikz}
\usepackage{tkz-graph}
\usetikzlibrary{arrows.meta,automata}
\usetikzlibrary{positioning}
\usetikzlibrary{shapes}
\usetikzlibrary{patterns.meta}
\usetikzlibrary{mindmap}

\newcommand{\vei}{vehicle}

\newcommand{\transp}{transportation}

\newcommand{\info}{information}

\newcommand{\vv}{C2C\xspace}
\newcommand{\vx}{C2X\xspace}

\newcommand{\rl}{RL\xspace}

\newcommand{\mas}{multiagent system\xspace}
\newcommand{\mass}{multiagent systems\xspace}
\newcommand{\marl}{MARL\xspace}

\newcommand{\ai}{AI\xspace}

\newcommand{\ar}{reinforcement learning\xspace}
\newcommand{\tap}{TAP\xspace}

\newcommand{\ue}{UE\xspace}

\colorlet{simul}{red!20!white}
\colorlet{info}{cyan!40}  %blue!40!white}
\colorlet{semaf}{yellow!40!white}
\colorlet{coadap}{green!20!white}
\colorlet{LightGray}{gray!20} %{rgb}{1,1,0.8} 
\colorlet{Gray1}{gray!70} %{rgb}{1,1,0.8} 
\colorlet{Gray2}{gray!90} %{rgb}{1,1,0.8} 
\definecolor{hublue}{rgb}{0.08,0.31,0.59}
\definecolor{hured}{RGB}{138,15,20}
\definecolor{hugreen}{RGB}{0,87,44}
\definecolor{husand}{RGB}{210,192,103}
\definecolor{hugraygreen}{RGB}{209,209,194}
\definecolor{hugrayblue}{RGB}{189,202,211}
\usetikzlibrary{%
	chains,
	calc,
	shapes,
	decorations,
	scopes,
	arrows,
	svg.path,
}

\tikzstyle{lane-right}=[start chain=going right,node distance=1ex]
\tikzstyle{lane-left}=[start chain=going left,node distance=1ex]

\tikzset{Car/.style={path picture={
\pgftransformscale{1em/10mm}
\fill[#1] svg "M28.21-13.605c9.562,3.446,9.648,23.671,0.155,27.315 c-4.363,0.507-11.834,0.663-14.102-0.052l-15.559,0.098c-0.361,0.54-2.344,0.587-2.49-0.046l-9.12-0.052 c-2.42,0.596-9.124,0.491-13.695,0.354c-0.868-0.069-2.466-0.383-3.055-0.86c-0.586-0.088-2.902-0.613-3.526-2.013 c-2.881-3.816-3.105-17.391-0.049-22.328c0.849-1.417,2.686-1.826,3.369-1.86c1.172-0.46,2.07-0.8,2.747-0.914 c4.707-0.249,12.097-0.404,14.099,0.304h9.073c0.314-0.542,2.336-0.684,2.805-0.049l15.296,0.103 C16.197-14.515,25.906-14.046,28.21-13.605z";%
\fill[CarWindow] svg "M-24.197,9.442c-0.964,0.555-2.364,1.198-2.94,1.826c7.256,0.867,22.892,1.016,38.82,1.092 c-1.101-0.841-5.213-1.443-8.749-2.128C-0.618,9.9-14.507,9.656-24.197,9.442z";%
\fill[CarWindow] svg "M33.182,7.616l-1.928,0.117c-0.877,1.522-1.91,4.131-4.51,5.058 c0.353,0.272,0.898,0.205,1.379,0.233C31.92,11.97,32.73,8.522,33.182,7.616z";%
\fill[CarWindow] svg "M-31.825-9.373c-2.358,4.111-2.324,14.41,0.069,18.753l4.252-0.791 c-2.718,0.159-2.64-17.187,0.057-17.171L-31.825-9.373z";%
\fill[CarWindow] svg "M14.187-11.443c-2.099,0.058-6.936,1.246-10.511,1.89C5.151-6.242,4.517,7.381,3.746,9.752 c3.689,0.599,8.887,1.858,10.498,1.517C18.698,9.763,18.805-9.979,14.187-11.443z";%
\fill[CarWindow] svg "M11.803-12.355c-15.93,0.087-31.488,0.225-38.756,1.103c0.581,0.619,1.916,1.206,2.886,1.758 c9.685-0.21,23.573-0.4,27.118-0.732C6.597-10.911,10.71-11.511,11.803-12.355z";%
\fill[CarWindow] svg "M27.243-13.147c-0.235,0.029-0.453,0.103-0.626,0.249c2.599,0.919,3.635,3.526,4.503,5.049 l1.934,0.127c-0.438-0.917-1.317-4.36-5.118-5.424C27.691-13.128,27.474-13.162,27.243-13.147z";%
},minimum width=2.5em,minimum height=1em},
Car/.default={black},
CarWindow/.style={fill=white},
}

\tikzset{Rsu/.style={path picture={
\pgftransformscale{3em/14mm}
\pgftransformrotate{180}
\fill[#1] svg "M5.188,18.438c-0.242,0-0.454-0.175-0.493-0.422L0.697-7.22C0.654-7.492,0.84-7.749,1.113-7.792
	c0.271-0.043,0.528,0.143,0.572,0.416l3.998,25.235c0.043,0.272-0.143,0.529-0.416,0.572C5.241,18.436,5.214,18.438,5.188,18.438z";
\fill[#1] svg "M-5.624,18.74c-0.026,0-0.053-0.002-0.08-0.006c-0.272-0.044-0.458-0.301-0.414-0.573L-1.682-9.34
	c0.043-0.272,0.299-0.461,0.573-0.414c0.272,0.044,0.458,0.301,0.414,0.573L-5.131,18.32C-5.17,18.565-5.383,18.74-5.624,18.74z";
\fill[#1] svg "M-5.019,17.469c-0.163,0-0.322-0.079-0.418-0.226c-0.151-0.231-0.087-0.541,0.144-0.692l9.001-5.911
	c0.23-0.151,0.541-0.087,0.692,0.144c0.151,0.231,0.087,0.541-0.144,0.692l-9.001,5.911C-4.83,17.442-4.925,17.469-5.019,17.469z";
\fill[#1] svg "M4.583,17.263c-0.094,0-0.189-0.026-0.273-0.082l-8.649-5.663c-0.231-0.151-0.295-0.461-0.145-0.692
	c0.152-0.23,0.461-0.295,0.692-0.145l8.649,5.663c0.231,0.151,0.295,0.461,0.145,0.692C4.906,17.183,4.746,17.263,4.583,17.263z";
\fill[#1] svg "M-3.774,10.612c-0.157,0-0.312-0.074-0.409-0.211c-0.159-0.226-0.105-0.538,0.12-0.697l6.807-4.806
	C2.97,4.74,3.282,4.793,3.441,5.018c0.159,0.226,0.105,0.538-0.12,0.697l-6.807,4.806C-3.574,10.583-3.674,10.612-3.774,10.612z";
\fill[#1] svg "M3.63,10.612c-0.1,0-0.2-0.03-0.288-0.092l-6.875-4.855c-0.226-0.159-0.279-0.471-0.12-0.697
	c0.159-0.226,0.471-0.279,0.697-0.12l6.875,4.855c0.226,0.159,0.279,0.471,0.12,0.697C3.941,10.539,3.787,10.612,3.63,10.612z";
\fill[#1] svg "M-2.935,4.747c-0.143,0-0.285-0.061-0.384-0.179c-0.177-0.212-0.149-0.527,0.062-0.704l5.195-4.346
	C2.152-0.66,2.466-0.63,2.643-0.419C2.82-0.208,2.792,0.108,2.58,0.285L-2.615,4.63C-2.708,4.709-2.822,4.747-2.935,4.747z";
\fill[#1] svg "M2.792,4.747c-0.113,0-0.227-0.038-0.32-0.116l-5.198-4.347C-2.938,0.107-2.966-0.208-2.79-0.42
	c0.177-0.21,0.492-0.24,0.705-0.063l5.198,4.347c0.212,0.177,0.24,0.493,0.063,0.705C3.077,4.686,2.935,4.747,2.792,4.747z";
\fill[#1] svg "M-0.005-12.703c-1.03,0.001-1.999,0.404-2.726,1.135C-3.459-10.84-3.858-9.87-3.856-8.827
	c0.005,2.123,1.735,3.851,3.857,3.851h0.006c1.03-0.002,1.998-0.405,2.726-1.135c0.727-0.73,1.126-1.699,1.125-2.741
	C3.853-10.976,2.123-12.703-0.005-12.703z";
\fill[#1] svg "M-4.533-3.745c-0.083,0-0.168-0.021-0.247-0.065c-1.347-0.765-2.315-2.009-2.726-3.502
	c-0.411-1.494-0.215-3.057,0.552-4.404c0.492-0.865,1.201-1.591,2.05-2.1c0.237-0.142,0.543-0.065,0.686,0.171
	c0.142,0.237,0.065,0.544-0.171,0.686c-0.702,0.421-1.288,1.021-1.695,1.738c-0.634,1.114-0.796,2.408-0.457,3.643
	C-6.202-6.342-5.4-5.313-4.286-4.68c0.24,0.136,0.324,0.441,0.188,0.682C-4.19-3.836-4.359-3.745-4.533-3.745z";
\fill[#1] svg "M-5.88-1.348c-0.083,0-0.168-0.021-0.246-0.065c-1.986-1.125-3.416-2.956-4.025-5.156c-0.609-2.2-0.325-4.505,0.799-6.492
	c0.723-1.274,1.77-2.35,3.025-3.109C-6.09-16.313-5.782-16.236-5.64-16c0.143,0.236,0.067,0.544-0.169,0.687
	c-1.11,0.671-2.034,1.621-2.673,2.747c-0.993,1.753-1.244,3.789-0.706,5.731c0.538,1.943,1.8,3.56,3.554,4.553
	c0.241,0.136,0.325,0.441,0.189,0.682C-5.536-1.439-5.706-1.348-5.88-1.348z";
\fill[#1] svg "M-7.574,1.059c-0.088,0-0.177-0.023-0.257-0.071c-2.627-1.581-4.482-4.089-5.223-7.065c-0.74-2.975-0.277-6.061,1.303-8.689
	c0.958-1.591,2.303-2.94,3.892-3.9c0.237-0.144,0.544-0.067,0.687,0.169c0.143,0.236,0.067,0.544-0.169,0.687
	c-1.45,0.877-2.679,2.108-3.553,3.561c-1.442,2.399-1.865,5.216-1.189,7.932c0.675,2.716,2.369,5.006,4.768,6.449
	c0.236,0.143,0.313,0.45,0.171,0.687C-7.239,0.972-7.405,1.059-7.574,1.059z";
\fill[#1] svg "M4.536-3.745c-0.174,0-0.343-0.091-0.435-0.253C3.964-4.238,4.048-4.543,4.289-4.68c1.113-0.633,1.914-1.662,2.253-2.898
	c0.34-1.235,0.178-2.53-0.456-3.644c-0.404-0.712-0.99-1.313-1.694-1.738c-0.237-0.143-0.313-0.45-0.17-0.686
	c0.142-0.238,0.45-0.314,0.686-0.17c0.851,0.513,1.56,1.239,2.048,2.101c0.766,1.346,0.961,2.91,0.551,4.403
	c-0.41,1.493-1.377,2.737-2.723,3.502C4.705-3.766,4.62-3.745,4.536-3.745z";
\fill[#1] svg "M5.882-1.348c-0.174,0-0.344-0.091-0.436-0.253C5.31-1.841,5.395-2.146,5.635-2.283c3.62-2.05,4.897-6.664,2.848-10.285
	c-0.64-1.127-1.564-2.077-2.674-2.746c-0.236-0.143-0.312-0.45-0.17-0.687c0.143-0.237,0.449-0.313,0.687-0.17
	c1.256,0.758,2.302,1.833,3.027,3.109c2.322,4.101,0.875,9.326-3.224,11.648C6.05-1.369,5.965-1.348,5.882-1.348z";
\fill[#1] svg "M7.576,1.059c-0.17,0-0.335-0.086-0.429-0.242C7.005,0.58,7.082,0.272,7.318,0.13c4.95-2.978,6.555-9.429,3.576-14.381
	c-0.873-1.451-2.101-2.683-3.55-3.561c-0.236-0.143-0.312-0.451-0.168-0.687c0.144-0.236,0.451-0.311,0.687-0.168
	c1.587,0.962,2.932,2.311,3.889,3.9c3.263,5.425,1.505,12.492-3.917,15.753C7.753,1.036,7.664,1.059,7.576,1.059z";
},minimum width=2.2em,minimum height=3em},
Rsu/.default=black}

\tikzset{Server/.style={path picture={
\pgftransformscale{3em/100cm}
\pgftransformyscale{-1}
\fill[#1] (-34.646,-26.568) -- (-34.648,36.998) -- (-5.069,42.52) -- (-5.069,-22.815) -- cycle;
\fill[white] (-18.734,37.15) -- (-21.861,36.579) -- (-21.861,7.092) -- (-18.734,7.43) -- cycle;
\fill[white] (-9.257,5.562) -- (-30.484,2.862) -- (-30.484,-3.595) -- (-9.257,-0.895) -- cycle;
\fill[white] (-9.257,-3.595) -- (-30.484,-6.296) -- (-30.484,-12.754) -- (-9.257,-10.053) -- cycle;
\fill[white] (-9.257,-12.755) -- (-30.484,-15.456) -- (-30.484,-21.914) -- (-9.257,-19.212) -- cycle;
\fill[#1] (-4.187,-24.749) -- (32.57,-39.392) -- (7.907,-42.52) -- (-32.506,-28.34) -- cycle;
\fill[#1] (-3.039,-23.021) -- (-3.039,42.166) -- (34.648,21.965) -- (34.648,-38.036) -- cycle;
},minimum width=3em,minimum height=3em},
Server/.default=black}

\graphicspath{{./home/bazzan/Images/}}

\title{Improving Urban Mobility: using artificial intelligence and new technologies to connect supply and demand
}

\author{
  Ana L. C. Bazzan \\
  Instituto de Inform\'atica, UFRGS, Caixa Postal 15064 \\91.501-970~~Porto Alegre, RS, Brazil \\ 
  \texttt{bazzan@inf.ufrgs.br} \\
}

\begin{document}
\maketitle 

\begin{abstract}
As the demand for mobility in our society seems to increase,
the various issues centered on urban mobility are among those that worry most city inhabitants in this planet.
For instance, how to go from A to B in an efficient (but also less stressful) way?
These questions and concerns have not changed even during the covid-19 pandemic; on the contrary, as the current stand, people who are avoiding public transportation are only contributing to an increase in the vehicular traffic.
The are of intelligent transportation systems (ITS) aims at investigating how to employ information and communication technologies to problems related to transportation. This may mean monitoring and managing the infrastructure (e.g., traffic roads, traffic signals, etc.). However, currently, ITS is also targeting the management of demand. In this panorama, artificial intelligence plays an important role, especially with the advances in machine learning that translates in the use of computational vision, connected and autonomous vehicles, agent-based simulation, among others.
In the present work, a survey of several works developed by our group are discussed in a holistic perspective, i.e., they cover not only the supply side (as commonly found in ITS works), but also the demand side, and, in an novel perspective, the integration of both.
\end{abstract}

\section{Introduction}

Despite climate change issues, it seems that our society is turning more and more mobile.
Covid-19 has its share of responsibility, as many people is avoiding public transit. While cycling may be an option in developed countries that already provide bike lanes and the regulation for a safety bike trip, this is not the case in developing countries, where the number of trips using private vehicles  increased even more during the pandemic. For obvious reasons, car sharing is not an option under such circumstances.
Even before the pandemic, there has been a growing demand for mobility. For example, according to  INRIX, congestion costs each American nearly 100 hours, which amounts to US\$1,400 per year\footnote{\url{https://inrix.com/press-releases/2019-traffic-scorecard-us/}}.
In 2017,  costs were of the order of 300 billion
dollars, an increase of 10 billion compared to 2016. 

The direct and indirect impact of
congestion in urban and interurban areas is immense and results in
costs that can reach up to 1\% of GNP.
According to experts, these costs are of various types. Among them, two are prevailing: 
opportunity cost and  monetary expenses imposed, not only on the driver(s), but also on society. Examples of these are
expenses related to the consumption of fuel, as well as expenses regarding health, caused by various kinds of  pollutants. 
Besides, the negative impact is also felt in the structure
country's economic growth, health, and quality of life.
``Solutions'' such as  tolls, license rotation etc., practiced
currently in Brazil,  are extremely unpopular. Citizens need
to see the return of their sacrifice, whether monetary or not. Thus,
there is a great demand for solutions that involve intelligence and
information as a way to offer a counterpart to the population. 

From a practical point of view, the question of how to efficiently move from A to B  is a topic that is on the agenda of most inhabitants
of the cities of the planet. This can be seen by the number of applications to assist  choosing a route to drive  or plan the use of public transportation means. 
A  way to mitigate traffic congestion  is to make better use of the existing infrastructure.
Fortunately, scientific and technological advances today allow us to
be optimistic about this task.

On the scientific side,
recent advances in artificial intelligence (AI) research facilitate  optimizing the use of the existing infrastructure.
This ranges from smarter traffic control to services that
not only indicate less congested routes to the users of the
transportation system, but they do so in order to balance the use of the road network as a whole. 
Such agenda also involves:  microscopic, agent-based simulation; computer vision (the basis not only for monitoring roads, but also to turn autonomous vehicles a reality); deep neural networks; and other techniques that emerged from AI.

On the technological side, AI has been associated with the use of solutions and products 
provided by the recent advances in the area of data networks, Internet and, more
specifically, IoT (Internet of Things).
The latter allows sensors to be embedded both in vehicles and in the road infrastructure, thus forming the basis for connected and autonomous vehicles.

It is in this context that the agenda around smart cities emerges,
where one focus is on smart urban mobility -- as for instance, the rational use of
different means of transport, integrating them and adapting them to demand. 

In more general terms, several researchers point to a scenario where the
Internet will be prevalent in vehicles and replace, at least in part, the Internet
as we know  today. For example, in ``Reinventing the Automobile'',
Mitchell and colleagues \cite{Mitchell+2010} claim that the so-called mobility
internet  will enable vehicles to have the same abilities that the computers now have regarding the standard Internet: exchange a huge amount of
geo-referenced information, in real time, which will allow integrating
vehicles to the IoT. This will potentially influence the way one manages
and optimizes travel on a road network. However, many of these
services have conflicting objectives when studied at the level of
component of the system. For example, it is known that a naive diffusion of
 information (e.g., same information for most of the drivers) can have negative consequences \cite{Wahle+2000,Wahle+2002}. 
 Further, even  an intelligent traffic signal management has to be carefully designed, as the performance
of each signal is strongly linked to  adjacent ones.

Although the scenario imagined by Mitchell and colleagues is not yet deployed,  it is already being employed in prototypes of various magnitudes in research labs, such as ours.
In fact, for the last two decades, the author has proposed, 
developed, and applied \ai techniques to various problems around the agenda of urban mobility.
One expectation is to add up to developing   public policies that lead to  more
smart cities. 
Quoting Martin Wachs,  ``\textit{Mobility and increased access to transportation are two of the most important global forces for the alleviation of poverty.}'' \cite{Wachs2010}.

As an example (others will be presented in Section~\ref{sec:met}), our laboratory has developed an open source traffic simulation infrastructure. 
Its advantage  becomes evident when one thinks that traffic is a complex and dynamic system. As such, the best way to investigate the emergence of patterns is by using simulation tools.
Indeed, there is an urgency  for acting on traffic systems due to: (i) climate change and need to reduced carbon emissions and (ii) the pandemic 
caused by the Sars-CoV-2 virus. If the adoption of  measures for reducing carbon emissions were already in the agenda,  the
pandemic shed light on the need for new policies. 
For instance,  Anne Hidalgo (mayor of Paris) proposed an important
change in the city by blocking private vehicular traffic in the
Rue de Rivoli, allowing only bicycle traffic in three
lanes, the fourth being reserved for buses and taxis. The mayor of New
York is  focusing on outdoor activities, 
temporarily excluding vehicular traffic over 110 kilometers of
streets, and allowing bicycles and outdoor tables for
restaurants. These examples just shed  light  on the need to
optimize existing resources in the city. 

Thus, the purpose of this text is to discuss the various works
carried out by the author. Such works aim both at optimizing the supply (for
example, with intelligent control of traffic signals), and at  distributing the
demand (with non-trivial dissemination of information and recommendations for 
travelers).

The rest of this text focus on  \ai's contribution and how it is likely to be even more prominent, especially considering  large volumes of data collected from 
personal mobile devices and/or embedded sensors. 
The vision is that there will be a truly intelligent system, with individuals, traffic signals and
vehicles connected and working together. In this view, traffic signals receive information about the state of the 
traffic at network level, in order to assess how recurrent and non-recurrent  events may influence the policies regarding signal timings. In addition, there will be an exchange of information among connected and autonomous vehicles,
improving their distribution in the transport network. Finally, from the
point of view of the system as a whole, the current research seeks a multiobjective reinforcement approach approach, where it aims not only at 
reducing travel times, but also reducing  emissions,
opportunity costs, and the effects of negative externalities.

\section{Theoretical Framework and State of Knowledge}
\label{sec:bkg}

The problem of how to move from A to B efficiently seems
increasingly complex and is among the main concerns of a
typical urban citizen. How did we get to this scenario? Putting it
simply, whenever the demand exceeds the supply, 
congestion arises. 
For the sake of  illustration, we can think of the following
decision scheme: 
In the past, when there were few users in the
road network, decisions were easily made and implemented,
resulting in rare or no congestion. 
One can say  that the (traffic) system had a low coupling. 
From the second half of the twentieth century this
situation has changed with an increase in the demand side, and, to a lesser extent,  in the supply side.
With such increase in complexity, it is expected that users no longer know the entire road network. 
This way,  the system's coupling  has increased significantly, causing many decisions to affect the decisions of  others. Hence, methods of traffic control have emerged.
With time, though,  their efficiency seems to be reducing. This is the phase we are currently experiencing. New
technologies aim at returning the system to the situation when users have  a higher level of
information.

This agenda is manifold and   multidisciplinary in its nature.
Thus, this section
briefly presented some fundamental concepts that underly
the works discussed in Section~\ref{sec:met}. 
More details on concepts about transport systems and traffic simulation can be
found in \cite{Bazzan&Kluegl2007,Bazzan&Kluegl2013book,Bazzan&Kluegl2014,Kluegl&Bazzan2012}. 

Sections~\ref{sec:mobits} and~\ref{sec:tap} address issues
directly related to demand: how
go from A to B, which involves urban mobility, intelligent
transportation, and the problem of how to allocate trips to the 
existing  infrastructure. 
Following, sections~\ref{sec:simul} to~\ref{sec:newtec}
discuss simulation techniques (necessary to study the effects of the
use of certain control measures), as well as
as concepts linked to \ai and the use of new technologies. In turn, 
sections~\ref{sec:utc} and~\ref{sec:atis}  discusses the
state-of-knowledge in the areas of traffic signal control and guided navigation, respectively. 
Section~\ref{sec:open} presents a summary of the open questions.

\subsection{Urban Mobility and Intelligent Transportation Systems}
\label{sec:mobits} 

Transportation and traffic experts have long been working with 
computational tools for demand estimation, as well as for optimizing the use  of the
infrastructure. 
The ITS (Intelligent Transportation Systems) area has a
multidisciplinary character and arose precisely for, among others
objectives, encourage the use of new technologies. 
Among these, in the last years, 
\ai has influenced the ITS area, by increasing 
 the performance of optimization and control processes. The goal of 
ITS is to develop  correct,  safe, scalable, persistent, and ubiquitous control systems. 
However, traffic control alone cannot solve the aforementioned problems. 
Control is just one facet of ITS; there are others
providing up-to-date and reliable information to the various
users of these systems, in a \textit{human-in-the-loop} manner.

As aforementioned, ITS involves the application of modern technologies from  the information and communication technology area. It can be said that ITS comprises
two major areas: ATMS (advanced travel management systems) and ATIS (advanced
traveler information system). 
While the first refers to the infrastructure and 
engineering side (supply), the second is directly related to the user of the
transportation system (demand).

An ATMS aims to safely and efficiently  manage technologies linked to
control devices, as well as to traffic monitoring, communication, and  control devices. 

An  ATIS aims at providing information to drivers and other users of the
transportation system. 
This \info\ is generated mostly  by an ATMS, eventually handled by the engineering team, and
then transmitted to users, by diverse means: from radio broadcast to private services developed for users who subscribe such a service or have dedicated,  embedded devices in their vehicles. These may   include a proper route guidance or, most commonly, just a panorama of the status of the traffic level on  the network.

\tikzstyle{block} = [rectangle, draw, fill=orange!50, 
		text centered, rounded corners,
		]
\tikzstyle{line} = [draw, -latex']

\begin{figure}
	\centering    
\begin{tikzpicture} [node distance = 2cm, auto]
    % Place nodes
    \node [block] (decision) {\normalsize travellers' decisions};
    \node [block, below of=decision, node distance=1.5cm] (demand) {demand};
    \node [block, right of=demand,  node distance=3cm] (routes) {routes};
    \node [block, below of=demand,  node distance=1.5cm, xshift=1cm] (perception) {\normalsize perception of traffic pattern};
    % Draw edges
    \draw [-{Latex[length=3mm]}] (decision) -- (demand);
    \draw [-{Latex[length=3mm]}] (demand) -- (routes);
    \draw [-{Latex[length=3mm]}] (routes.east)  |-  (perception.east);
    \draw [-{Latex[length=3mm]}] (perception.west) -| (-2.5,0) -- (decision.west);
    \draw [-{Latex[length=3mm]}] (perception) -- (routes);
\end{tikzpicture}
\caption{Feedback scheme: decision, demand, route, and perception.}
\label{fig:feedback}
\end{figure}
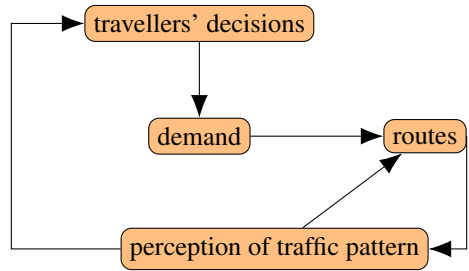

These technologies can potentially help  mitigating the 
effects of the growing demand for mobility. However, there is a
well-known feedback phenomenon: the decisions of the
travelers generate a demand, which translate into route choices, 
which in turn create  a traffic pattern. Travelers may then  review their decisions (or, directly, their choices of
routes) and the cycle is repeated (see Figure~\ref{fig:feedback}). 
This is a complex process that requires that any  measure is first 
tested  in a simulation environment. 
For this, there are classic simulation methods for traffic engineering that
are able to model both the infrastructure and the demand with different degrees of
fidelity. However, some of the new technologies related to an 
ITS, such as ATIS, are not trivial to simulate, as they involve complex human behaviors.

\subsection{The Traffic Assignment Problem}
\label{sec:tap}

In a  \transp\ network, travellers' trips are distributed among   a set of origin-destination
(OD)  pairs. Each of these pairs is associated with  several paths or routes that connect its respective origin and destination. 
The   traffic assignment problem (\tap) consists of
assigning trips to such routes in an optimal way, given restrictions of
capacity and others. 

In general, each user knows the best route 
between two points, under the assumption that  the links in the network are not congested. 
However, this is not realistic.
In situations such as peak hours, the traffic pattern changes and routes that were not optimal before may become
attractive alternatives. Commuters, who are familiar with the conditions of the
network,  tend to perform an 
individual optimization  process, based on their own experience. 
In a situation in which each traveller has found the route that has the shortest travel  time, none has an incentive to change its choice. This  state is called user  (or Nash) equilibrium, as formulated by
Wardrop \cite{Wardrop1952}: no user can improve his performance
by changing routes, which means  that all routes have equal costs (Wardrop’s First Principle). 
This state is maintained if  neither the demand nor the network changes.

Although it appears to be an interesting result, two questions are
relevant. First, variations in demand conditions, 
special events (sports, cultural), or weather  can
change traffic conditions. The equilibrium state can be dissolved, leading, for example 
to highly congested arterials, while other parts of the network
are underused. 
It is in this circumstance that an ATIS can be used
to improve the efficiency at network level, via direct or indirect recommendation of 
alternative routes. 

A second problem is that the user equilibrium (henceforth \ue), as formulated by Wardrop's First Principle, 
is not necessarily efficient from the point of view of the system as a whole because, in general, it leads to a greater sum of the travel time. This stems from the fact that a myopic or greedy route choice can cause an increase in the travel time and the consequent degrade in performance of the network as a whole. 
This way, an ATIS can be used,  so that the traffic manager can align the optimum of the system
(the desired situation from the point of view of the system) with the 
\ue (a not necessarily efficient situation). 
In a more recent view,
this problem has been formulated as a congestion game, thus using  game theory techniques
\cite{Roughgarden&Tardos2002}.

\subsection{Traffic Simulation}
\label{sec:simul}

Transport systems are complex systems, whose behavior is
defined by interactions between various types of entities. 
As tests in the real-world are costly, time-consuming, or may pose safety issues, simulation tools are widely used.

In transportation, simulation models are usually classified according to the level
of detail with which they represent the  system. These levels
are: macroscopic, mesoscopic and microscopic. A microscopic model
describes the  entities and their interactions at a high level of
detail. For example, in terms of movement, it is possible to model vehicles changing lanes  and car-following  behavior. A mesoscopic model 
generally represents the majority of the entities at a more reasonable and efficient level of detail, modeling the   interactions between those entities at a higher level
of abstraction. A macroscopic model describes both the entities
and the interactions between them at a high level of abstraction. For 
example, a macroscopic model represents traffic by means of a set of variables that aggregate
values from histograms of flows, densities and
speeds. A lane change maneuver is not even represented,
since \vei s do not exist as separate entities; rather, they are
represented only as density aggregates. However, 
macroscopic models have the advantage of being computationally efficient and are useful for predictions 
that regard  density, speed and flow. In general these models are highly
sensitive to the initial parameters, which can have an  impact
on the overall behavior of the system. On their turn, microscopic models are generally complex and costly to develop.
In addition, they  involve a much larger number of parameters and thus require more data.
Yet, given the need to model individual behaviors, the trend is toward use of microscopic models, especially in tasks that involve simulating effects of traffic control measures and/or involve human behavior.

At the microscopic modeling level, there has been a growing interest in the use of
\ai techniques. One of the most interesting paradigms is that of
agent-based simulation (ABS). This if
due to the fact that, in ABS, it is possible to model an individual driver's decision-making, 
increasing the possibilities of modeling heterogeneity and
individuality (for example for route recommendations).

\subsection{Distributed AI}
\label{sec:dai}

This section briefly introduces concepts about distributed
\ai  and \mass; for details see \cite{Wooldridge2009,Bazzan2010}.

Up to the 1970s,  \ai has focused on problem solving techniques involving a single entity (a robot,  a  vehicle, an expert, and so on). However, strictly speaking, none of these entities
exists in isolation. Moreover, these individuals may have complex interrelationships and coupling. 
This means that, in real-world applications, it might be expensive to consider such complex systems in their entirety and/or use a centralized modeling approach.
On the other hand, when decomposing a given problem, 
the interactions among the components  must be
carefully handled. This was the initial motivation that led to the arising of a new subarea of \ai, initially called 
distributed artificial intelligence (DAI).
Later, the focus of DAI has turned to \mass,  a view of DAI in which the components are not necessarily collaborative or have a common goal.

A   \mas is a system that consists of a number of agents that
interact with each other. One of the motivations for development
of \mass is the possibility to solve problems that
cannot be treated in a centralized way, due to issues such as  limitation of computation and/or communication resources, performance, or privacy.

\subsection{Reinforcement Learning}
\label{sec:rl}

Machine learning techniques have found more and more applications
in transportation. In particular, \ar is one of the most used techniques because it allows  different
classes of agents (for example traffic signal controllers, vehicles) to 
adapt to the state of the traffic, by building a model that tells 
the agent what action to do in each observed state. In this way, the
system designer does not need to provide the agent with models that require domain 
knowledge  or training instances that are of
difficult to obtain. 

There are two major classes of the \ar techniques:
model-based and independent of model. In the former, the
agent has  models that explain how the environment 
behaves (possibly under different conditions), as well as which reward to expect when selecting a given action in a given state. 
In the latter, the agent has to
construct such  models.
In both cases, agents aim at  maximizing their expected rewards.

%%%%%%%%% JAI
\subsection{New Technologies}
\label{sec:newtec}

In the last two decades, interest in new technologies has grown,
especially those linked to autonomous vehicles, where \ai has a
fundamental role. In parallel, advances in the areas of communication and
Internet now allow the  so-called autonomous and connected vehicles
(CAVs).

In terms of communication, autonomous vehicles can be equipped with systems, called \vv (car to car
communication) or C2I (car to infrastructure). To designate both,
\vx is used. With this, they can exchange information about the
traffic flow  conditions.

It is expected that autonomous vehicles   reduce the
emissions and lead to more safety as they ensure, for instance, that  lane changes are done more carefully. 
It is worth mentioning that in 1994, 50.4\% of accidents in
American interstate highways was caused by drivers errors,
among which 27.6\% occurred during lane changes.

\subsection{Related Work: traffic control} 
\label{sec:utc}

Traffic control techniques have existed for several decades and
derive mainly from  the areas of operations research  and 
control. 
A classical approach is the synchronization of
traffic signals,  so that vehicles can
cross an arterial  in one direction, with a constant speed and 
without stops (the so-called “green wave”). Well 
known   (generally commercial) softwares in this category are 
 TRANSYT \cite{Robertson1969,Transyt1988},
 SCOOT~\cite{Hunt+1981,Robertson&Bretherton1991}, 
 SCATS  \cite{Lowrie1982},
and   TUC \cite{Diakaki+2003}.

More recently, \ai and \mass techniques
have been employed, especially in connection with \ar.
In fact, the problem of traffic control can  be addressed from the point of view of \ar. 
In these cases, \ar is used by traffic signals to learn a
policy that maps states (usually queues at intersections)
to actions. 
Due to the number of works that employ \ar for 
traffic control, the reader is referred to  surveys  \cite{Bazzan2009,Mannion+2016,Wei+2021,Wei+2021,Yau+2017}. 

It is worth mentioning that  few studies employ
\ar simultaneously  at traffic signal control and for  route choice, as our
work (see Section~\ref{sec:coadap}).
In fact, this integration, as obvious as it seems, has received little
attention in the literature. In the work of \cite{Wiering2000}, drivers and
traffic signals learn simultaneously. Traffic signal controllers receive 
specific information about drivers' routes (for example, the
destination) to calculate the expected waiting time. This however, can be considered a  strong assumption. In addition, the underlying model in \cite{Wiering2000} is not entirely microscopic. The work of  \cite{Taale+2015} does not use
reinforcement learning, but, rather, a strategy based on  back-pressure 
to integrate traffic signals and  route choice. 
% The proposed approach was tested using only a macroscopic simulation model.  

\subsection{Related Work: route guidance}
\label{sec:atis}

Understanding how a driver select  routes is
key in an ATIS. Some milestones  in this
area are  \cite{Adler&Blue1998,Ben-Akiva+1991,Bonsall1992,Mahmassani&Chen1991} among others. 
However, in these, the travellers' response to such systems is not
considered. This is only possible when one employs a 
microscopic, agent-based simulation, as described in
Section~\ref{sec:simul}. 

As previously mentioned
(Section~\ref{sec:tap}), it is essential to consider both  the global and individual costs. One way to do this is through tolling mechanisms devised to  penalize some users, such as those using 
roads with higher traffic. In this sense, the authors in ~\cite{Sharon+2017} use adaptive tolls to optimize the chosen routes, as the toll fees  change. The authors focus on the optimum of the system, which
can be achieved by imposing costs on drivers. Similarly,  \cite{Buriol+2010}  
deals with the \tap in a 
centralized perspective to find an assignment by imposing tolls on some parts of the network. 

To deal with fluctuations in the flow of vehicles, the so-called
congestion tolls can improve the 
efficiency of the network and a better 
traffic distribution  \cite{Arnott+1990,Kobayashi&Do2005}.

However, tolls are unpopular and, in general, unfair. The work described in 
Section~\ref{sec:routing} aims at achieving similar results through the 
dissemination of information to users. There are not many works that
consider \ai in this context. Neural networks were used by \cite{Dia&Panwai2014} in order to
predict the route choices. However, the authors focus 
only on the impact of the messages and not on the impact of the
traffic distribution and travel time. Neural networks were also
used in \cite{Barthelemy&Carletti2017}. The network parameters are determined in a 
preliminary training. The output of the neural network is the action to be selected by the agent: stay or modify the route. The work of   \cite{Dias+2014} uses an
ant colony optimization algorithm. The difference is that, instead of using the
pheromone to attract ants, it aims at 
repelling them. The approach proposed by \cite{Claes+2011}  is also based on
ant colony optimization, combined with traffic prediction.
However, agents here also have centralized information.

\subsection{Summary of Open Questions}
\label{sec:open}

% TALVEZ COLOCAR AQUI A FIG. DO SLIDE GRU-BERRINI? e dizer que foi (parcialmente) resolvido com ... refs secoes

This section provided a brief dive into key topics, as well as an overview on related work.
Clearly, there are several open questions in the literature; these are organized by topics:

\begin{enumerate}
	\item  \tap and route guidance: regarding the \tap (Section~\ref{sec:tap}), recall  that the \ue 
    may  be inefficient. In fact, from the point of view of the whole system, it would be
    better to reach the so-called  system optimum (Wardrop’s second principle), where the mean or sum of  all travel times are minimal. This state      can only be achieved if the negative externalities caused by
    individual user choices are addressed in some way,
    such as tolls or route recommendations that lead to the  
    system optimum. One example is  our work  reported in the Section~\ref{sec:rl4routing}, regarding aligning the system optimum and the \ue.
    On the route guidance front, as mentioned,  for the sake of simplifying the modeling, it is usual to  treat both the vehicle and the driver as a
    single and homogeneous entity. This, however, does not consider the individual goals,
    preferences and intentions, although these have a    decisive  influence on the decisions. 
    In Section~\ref{sec:routing}, some  contributions in this direction are discussed.

\item Simulation: although microscopic simulation platforms are
    key (especially when it refers to investigating the effect of  traffic control measures),  this kind of platform is not fully used. One reason is that    most microscopic simulators are  commercial and expensive products. Section~\ref{sec:itsumo} thus discusses a  simulation platform     developed by us under  the open source philosophy.

\item  Traffic signal control: although there are several approaches to traffic signal 
    control, our work focuses on the application of    \ar, in order to tackle  scenarios that change their
    dynamics. The challenges here are multiple.

First, the    formulation and modeling  relates to multiagent \ar
    (\marl), that is, where the presence of more than one agent makes the problem
    more complex 
 \cite{Guestrin+2002,Hu&Wellman1998,Melo&Ribeiro2010,Stone2007,Zhang+2008}.
The fact that there are several agents
    learning simultaneously in the same environment makes the problem
    inherently non-stationary, as several agents are making changes in 
    the environment, while others try to learn. In addition, there are
    only a few works that consider co-learning between two
    classes of agents, such as traffic signals and drivers, an issue that is
    covered  in Section~\ref{sec:coadap}.

 A second issue is that \ar requires the deployment of sensors of various  types, something that can be expensive. In this sense, our work     (Section~\ref{sec:c2x4nav})  calls for using 
    connected vehicles and \vx\ as a way of collecting information, besides the use of  traditional sensors already used in traffic engineering.
%     ; recall that  vehicles themselves could act as sensors.

   Finally, an important issue is the fact that, in the \ar literature,
    there are only a few studies that deal with the issue of learning and decision-making 
    based on multiple objectives. In fact, the vast majority of
    works dealing with \mass in general and with
    \marl in particular, consider that agents seek to formulate
    policies that relate to a single objective (for example,  agents
    have objective functions that considers only travel time).
    However, most real-world problems involve more than one
    objective to be considered. This problem starts to get more
    attention      \cite{Radulescu+2020,Hayes+2021}.
    However, in many cases, the solution
    found is to use scalarization over the multiple
    objectives, creating a single objective function, which may not
    find all Pareto-efficient solutions. These questions are
    discussed in Section~\ref{sec:ongoing}, where we present our ongoing work.

\end{enumerate}

\section{Using \ai\ to mitigate traffic congestion}
\label{sec:met}

\begin{figure*}[t]
	
\begin{center}
\resizebox{\textwidth}{!}{
\begin{tikzpicture}[mindmap, grow cyclic, every node/.style= concept, concept color=orange!40, 
% 	level 1/.append style={level distance=5cm,sibling angle=90},
	level 1/.append style={level distance=5cm,sibling angle=72},
	level 2/.append style={level distance=3cm,sibling angle=45},
]
\node{Research \.Contributions}
child [concept color=simul] { node {\textbf{\ref{sec:itsumo}} Simulator} 
}
child [concept color=Tan] { node {\textbf{\ref{sec:ongoing}}\\  Multiobjective Decision} 
}
child [concept color=info] { node {\textbf{\ref{sec:routing}} Guided Navigation} 
	child { node {\textit{human-in-the-loop}}}
	child { node {BDI Modeling}}
	child { node {Game Theory and Route Choice}}
	child { node {\rl\ for Route Choice}}
	child { node {C2X and Route Choice}}
}
child [concept color=coadap] { node {\textbf{\ref{sec:coadap}} Co-Learning} 
}
child [concept color=semaf] { node {\textbf{\ref{sec:control}} Traffic Signal Control}
	child { node {Hierarchical \rl}}
	child { node {Model-Based \rl}}
	child { node {Constraint Optimization}}
	child { node {Swarm Intelligence}}
	child { node {Game Theory}}
}
;
\end{tikzpicture}
}
\end{center}
\caption{Content Map of Section~\ref{sec:met}.}
\label{fig:map}
\end{figure*}
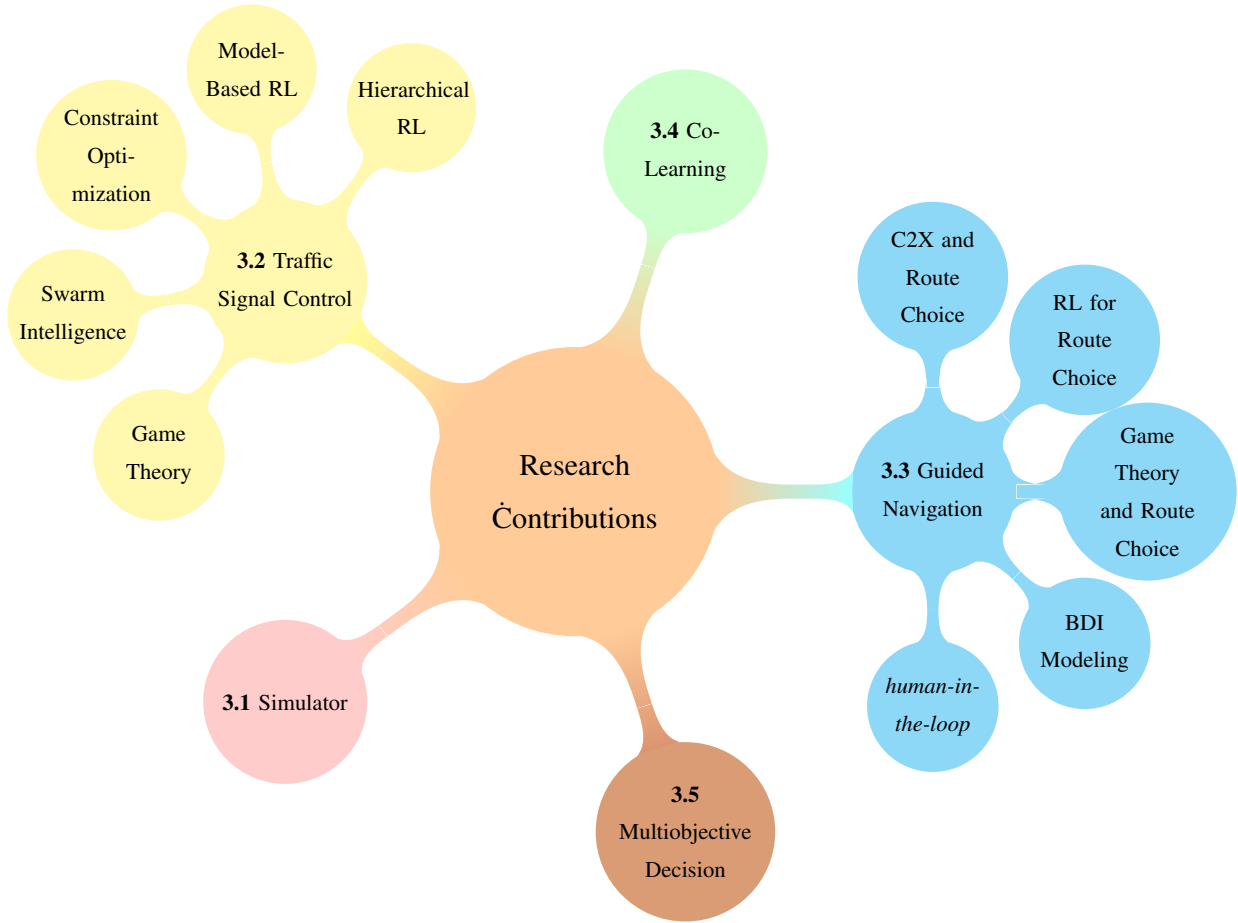

%Since the 1990s, the author has worked in the area of vehicular traffic, as 
% application domain for \ai techniques in general and DAI and \mass in particular, being recognized as a reference in the
% area, and having published a book () and articles of the type \textit{surveys} (\cite{Bazzan2009,Bazzan&Kluegl2014}).

This section discusses the main methods
proposed and developed by the author. Section~\ref{sec:itsumo} briefly discusses  the technological infrastructure developed for implementing  agent-based simulation. Following,  sections~\ref{sec:control} and~\ref{sec:routing} describe the main methods  that deal with the supply and demand sides respectively. Supply covers mostly intelligent traffic signal control, whereas demand refers to   guided navigation and route choice by the users of the
road network, which could be drivers or autonomous vehicles. 
Most of the employed methods use \ai in form of \ar. 

Given that in the real world both traffic signal controllers and road users learn simultaneously, this leads to  co-learning, which is much more challenging. 
Section~\ref{sec:coadap} then discusses works that address this scenario. 

Lastly, Section~\ref{sec:ongoing} reports work in progress, with emphasis for the topic of multiobjective \ar.

The map of the contents presented in this section - whose color scheme is
consistent throughout the text - is shown in Figure~\ref{fig:map}.

% \newpage
\subsection{Open Source Simulator}
\label{sec:itsumo}

\noindent
\resizebox{\textwidth}{!}{
\begin{tikzpicture}
\draw (-1,0) [fill=simul,draw=white] rectangle (14,2);
\node (sp2) at (0,1)
    {\includegraphics[width=0.12\textwidth]{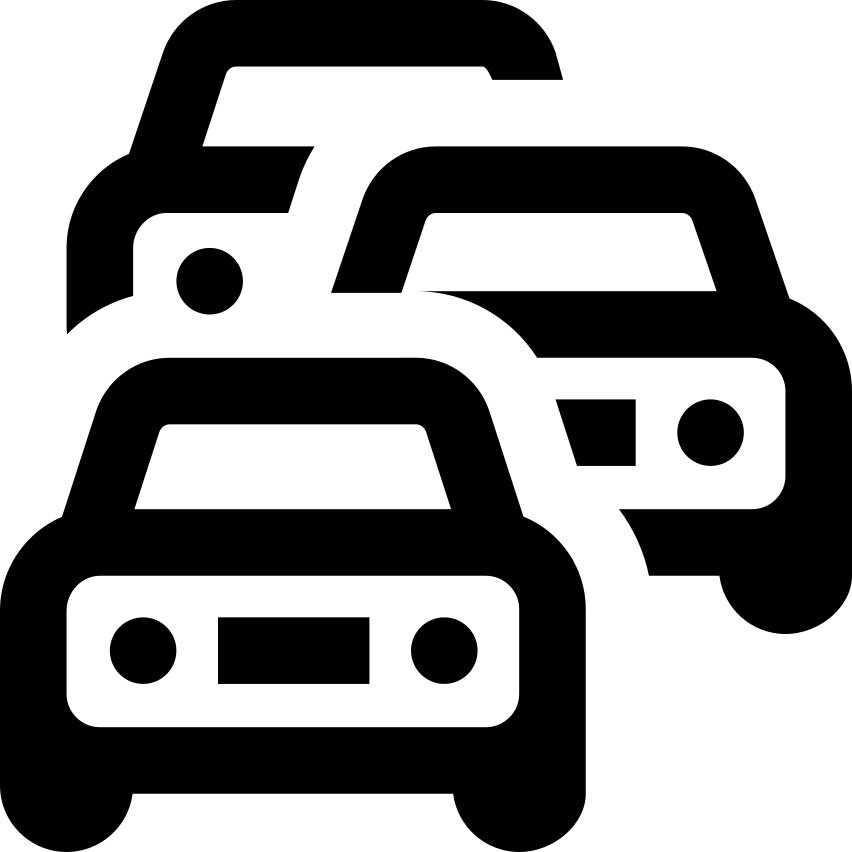}};
\node[anchor=west] (real) at (1,1)
    {\includegraphics[width=.11\textwidth]{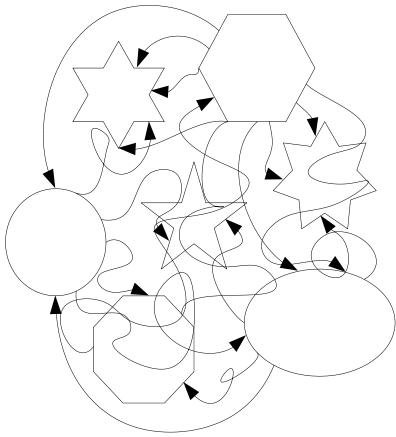}};
\node[anchor=center] (method) at (6.5,1)
%     {\includegraphics[width=.1\textwidth]{hammer.png}};
{\resizebox{.17\textwidth}{!}{ %% code to generate gears (source: https://tex.stackexchange.com/questions/58702/creating-gears-in-tikz)

% % #1 number of teeths
% % #2 radius intern
% % #3 radius extern
% % #4 angle from start to end of the first arc
% % #5 angle to decale the second arc from the first 
% 
% \newcommand{\gear}[5]{%
% \foreach \i in {1,...,#1} {%
%   [rotate=(\i-1)*360/#1]  (0:#2)  arc (0:#4:#2) {[rounded corners=1.5pt]
%              -- (#4+#5:#3)  arc (#4+#5:360/#1-#5:#3)} --  (360/#1:#2)
% }}  

% \begin{tikzpicture}
%    \draw[thick] \gear{18}{2}{2.4}{10}{2};
%  \end{tikzpicture} 
 
%%%%%%%%%%%%%%%%%%  another: 
% #1 number of teeths
% #2 radius intern
% #3 radius extern
% #4 angle from start to end of the first arc
% #5 angle to decale the second arc from the first
% #6 inner radius to cut off
% 
\newcommand{\gear}[6]{%
  (0:#2)
  \foreach \i [evaluate=\i as \n using {\i-1)*360/#1}] in {1,...,#1}{%
    arc (\n:\n+#4:#2) {[rounded corners=1.5pt] -- (\n+#4+#5:#3)
    arc (\n+#4+#5:\n+360/#1-#5:#3)} --  (\n+360/#1:#2)
  }%
  (0,0) circle[radius=#6] 
}

 \begin{tikzpicture}
	\coordinate (origin) at (0,0); 
   \fill[red, even odd rule]  \gear{18}{2}{2.4}{10}{2}{1};
   \begin{scope}[shift={($(origin.east)+(5,0.7)$)}]
    \fill[blue, even odd rule]  \gear{15}{1.8}{2.}{10}{2}{1};
    \end{scope}
   \begin{scope}[shift={($(origin.east)+(2.7,-2.9)$)}]
    \fill[black, even odd rule]  \gear{16}{1.5}{1.8}{30}{2}{0.7};
    \end{scope}
 \end{tikzpicture}
}};
\node[anchor=east] (model) at (11,1)
    {\includegraphics[width=.11\textwidth]{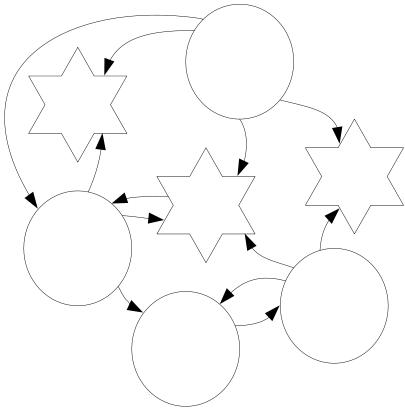}};
\node (cells) at (12.5,1)    
    {\includegraphics[width=.2\textwidth]{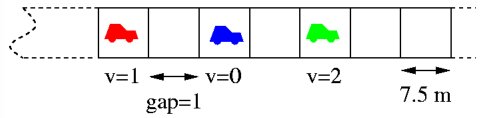}};
\draw[<->,thick,double] (real) -- (method) node at (4.2,0.7){real world to ...} ;
\draw[<->,thick,double] (method) -- (model) node at (8.3,0.7) {... model};
\end{tikzpicture}
}

As mentioned in Section~\ref{sec:simul}, simulation is heavily employed in traffic and transportation engineering. However, traffic simulation tools are complex and, mostly, available as commercial,  closed software. In general, one does not  have access to
source code; some commercial simulators do provide APIs that facilitate the interaction with the simulation, to different extents. Still, these tools tend to be expensive.
To mitigate this availability issue, our laboratory has built, over 10 years, an open tool to allow traffic simulation at 
microscopic level, called ITSUMO, which pre-dates SUMO \cite{Lopez+2018}.
Whereas SUMO is based on car-following,  which may be computationally demanding, the development of ITSUMO was centered on keeping computational costs low. Therefore, 
ITSUMO is based on a combination of an agent-based model
with the Nagel and Schreckenberg cellular automaton model \cite{Nagel&Schreckenberg1992}. While
the latter is computationally efficient because it is discrete in time and
in space, the agent paradigm allows great flexibility for
modelling the behavior of the various entities of the system, from the
traffic signal controller to the drivers, including their decision strategies.
Further, while one of SUMO's strength is the demand modeling, ITSUMO has focused on traffic signal control.

ITSUMO was the  basis that allowed the development of some of the works
mentioned ahead, in which
several innovative methods related to \ai and \ar have been proposed, tested and
applied to problems related to traffic control. 
% Figure~\ref{fig:itsumo} shows the ITSUMO modules (right) and applications (left) and the relationship between data entry, the simulator kernel, and the output provided.
Details can be found in  \cite{Bazzan+2010atss,Silva+2005}.

% \begin{figure}
% 	
% \begin{minipage}[left]{0.47\textwidth}
% 	\includegraphics[width=0.9\textwidth]{/traffic/ITSUMO_overview_new.png}
% \end{minipage}
% \hfill\vline\hfill
% \begin{minipage}[rigth]{0.47\textwidth}
% 	\includegraphics[width=0.9\textwidth]{/traffic/itsumo_scheme.jpeg}
% \end{minipage}
% \caption{ITSUMO Modules (left); input-output scheme, simulator kernel, output (right).}
% \label{fig:itsumo}
% \end{figure}
%  
% \vspace{0.5cm}
% \begin{minipage}[t]{0.95\linewidth} 
% \begin{tcolorbox}[colback=simul,colframe=red,title=Main result]
% Development of an agent- and CA-based microscopic traffic simulator.
% \end{tcolorbox} 
% \end{minipage}  

\subsection{Intelligent Traffic Control Using Machine Learning and Other AI Techniques}
\label{sec:control}

\noindent
\resizebox{\textwidth}{!}{
\begin{tikzpicture}
\draw (-1,0) [fill=semaf,draw=white] rectangle (14,2);
\node (dice) at (0,1)
    {\includegraphics[width=0.12\textwidth]{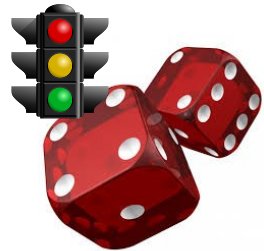}};
\node[anchor=west] (wave) at (3.4,1)
    {\includegraphics[width=.19\textwidth]{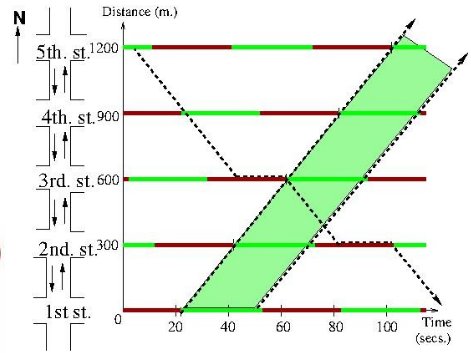}};
\node[anchor=center] (nash) at (7.5,1)
    {\includegraphics[width=.19\textwidth]{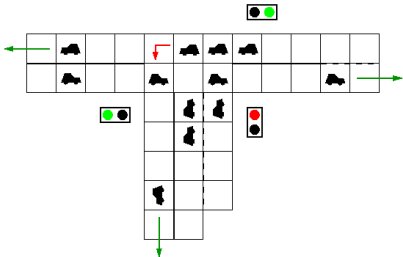}};
\node[anchor=east] (marl) at (14,1)
	{\begin{tikzpicture} 
		\draw[fill=gray!30,draw=white] (10,4) -- (10.5,4) -- (11,2.5) -- (10.5,2.5)  -- cycle;
		\draw[color=black, style=dashed] (10.25,4) -- (10.75,2.5); 
		\node at (11,4) {\includegraphics[width=.03\textwidth]{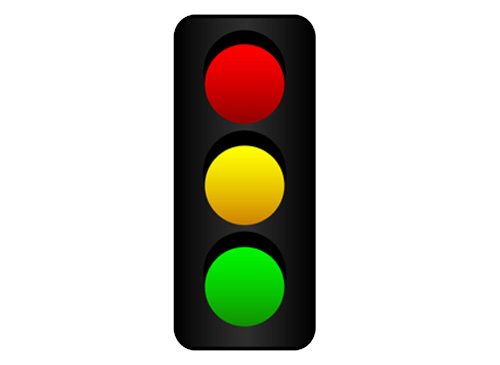}};
		\node at (11.2,3.8) {\includegraphics[width=.04\textwidth]{figs/signal}};
		\node at (11.5,3.5) {\includegraphics[width=.06\textwidth]{figs/signal}};
		\node (bigger) at (12.2,3.1) {\includegraphics[width=.1\textwidth]{figs/signal}};
		\node (s) at (13,3.5) {\scriptsize state?};
		\node at (13,3.0) {\scriptsize action: $a$};
		\node at (13,2.5) {\scriptsize reward: $r$};
		\draw[<->] (bigger.north) to [out=30,in=90]   (s.west);
	\end{tikzpicture}};
\end{tikzpicture}
}

The classic approaches described in Section~\ref{sec:utc} has some
disadvantages. In an attempt to address them, the following describes the
main approaches proposed and developed by the author.

\subsubsection{Traffic Signal Coordination Via Game Theory}

In \cite{Bazzan2005}, the author proposed the first approach to
formation of green waves where each traffic signal is modeled as an agent
that learns. Each has
pre-defined plans for synchronization / coordination with agents
adjacent in different directions according to the traffic situation.
This approach employed evolutionary game theory techniques, having the following benefits: agents can create subgroups of
synchronization to form green waves and prioritize the flow in a given 
direction; there is no need for central control; and there is neither communication
nor direct negotiation between agents. The approach was tested in a
arterial, obtaining better performance than a centralized approach
classic.

\subsubsection{Swarm Intelligence Based Approach}

In  \cite{Oliveira+2004} an approach to dynamic formation of
green waves  was presented, an an attempt to address some disadvantages of both
classic approaches and also those that are 
fully decentralized. This new approach is based on
swarm intelligence. Each traffic signal behaves like an insect
where the decision-making process is inspired by the task specialization process. It also uses the 
pheromone metaphor, i.e., it assumes that vehicles leave traces where they travel. However, pheromone here is used to repel other vehicles, rather than to attract others.

The approach was tested with an arterial route located in the city of
Porto Alegre.  It has been shown that the system tends to remain stable and
adapted to the traffic, without losing its ability to 
adapt to  changes in the environment.  

\subsubsection{Constraint Optimization Based Approach}
The approach described in \cite{Oliveira+2005}, unlike those in \cite{Bazzan2005} and  \cite{Oliveira+2004}, is based on explicit communication between agents,
as well as in cooperative mediation. In this model, the problem of forming
green waves was approached as a compromise between a coordination with only implicit communication  (\cite{Oliveira+2004}), and a classic centralized solution (Section~\ref{sec:utc}). 
An algorithm for distributed constraint optimization was used, where a mediator
helps to assign values to  decision variables. The scenario used for testing
extrapolates an arterial route, being a grid and therefore having a
high number of constraints. In the experiments, two
coordination directions were considered: horizontal and vertical. The mediation-based approach computes
more appropriate and flexible coordination groups, as compared to groups found by  classical methods.

\subsubsection{Model Based Reinforcement Learning Approach}

When dealing with non-stationary environments, where vehicle flow is not
constant, both model-independent \ar approaches and
model-based ones (see Section~\ref{sec:rl}) have drawbacks.
Specifically, when the environment changes, they both need to relearn everything
from scratch, since the policy calculated for a given environment
are no longer valid if there is a change in vehicle flow dynamics.
This causes \ar algorithms to experience performance drops. Further, these algorithms potentially have to relearn policies even for situations that have already been previously
experienced.

As discussed before, model-based approaches for \ar assume the existence of a fixed number of models that supposedly describe the 
behavior of an environment. Since this assumption is not always
realistic, an alternative is the incremental construction of models.

In the specific case of traffic signal control, a such method  was proposed in
\cite{Silva+2006icml}, 
where a
traffic signal controller is able to  automatically handle different flow dynamics. An optimal policy  is associated with each model, i.e., there is  a mapping
between the traffic conditions and the corresponding traffic signal plan to be
chosen.
This way, experiments have shown that performance drops can be avoided as there is no need for constant relearning from scratch, thus opening the way for the use of model-based approaches in traffic signal optimization.

\subsubsection{An Approach Based on Hierarchical Control}

The presence of  many signal controllers  present a challenge for \ar methods. To deal with
large-scale road networks that may have a high number of learning controller agents, some approaches have been proposed by the author.

In \cite{Bazzan+2010eaai} an approach has been proposed that considers that these agents are organized in
groups, each under the supervision of a manager agent. This manager thus controls a group containing several Intersections, computing and recommending joint actions, as opposed to actions that are locally computed and selected by the agents at the intersections. These, in turn, try to balance the actions recommended by the manager with actions that are optimal from their myopic and local point of view. 

In a similar direction,
\cite{Abdoos+2013eaai} proposes the
use of a holonic multi-agent system to model a road network
partitioned into regions (holons). The differential of this method was the
extension of the Q-learning method \cite{Watkins&Dayan1992} to the region level. 

In \cite{Abdoos&Bazzan2021}, 
a hierarchical multi-agent system is proposed, which includes two levels of traffic signals. In the first level, each signal is controlled by an agent. For the other levels, the traffic network is divided into a number of regions, each controlled by a region agent that controls the group. First level agents employ reinforcement learning to find the best policy, and then send their local information to the agent in charge of a region. Moreover, the local information is used to train a long short-term memory (LSTM) neural network for traffic status prediction. The agents in the above level can control the traffic signals by finding the best joint policy using the predicted traffic information. Experimental results show the effectiveness of the proposed method in a traffic network including 16 intersections.

In all these works, a significant acceleration in the
learning was achieved, which is to say that traffic signals adapt a lot
more quickly to new flow conditions.

\subsubsection{Traffic Control: main result}

\begin{minipage}[t]{0.95\linewidth} 
\begin{tcolorbox}[colback=semaf,colframe=Goldenrod] %,title=Principal resultado]
Development of several high-performance methods for adaptive and decentralized traffic control. 
\end{tcolorbox} 
\end{minipage}  

\subsection{Guided Navigation}
\label{sec:routing}

\noindent
\resizebox{\textwidth}{!}{
\begin{tikzpicture}
\draw (-1,0) [fill=info,draw=white] rectangle (14,2);
\node (where2) at (0.3,1)
    {\includegraphics[width=0.15\textwidth]{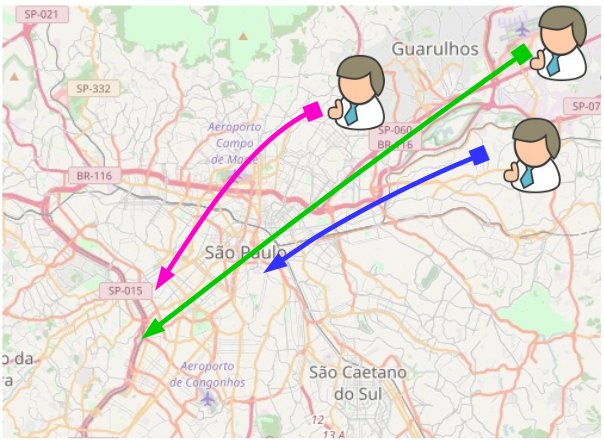}};
\node[anchor=west] (tap) at (1.5,1)
    {\includegraphics[width=.32\textwidth]{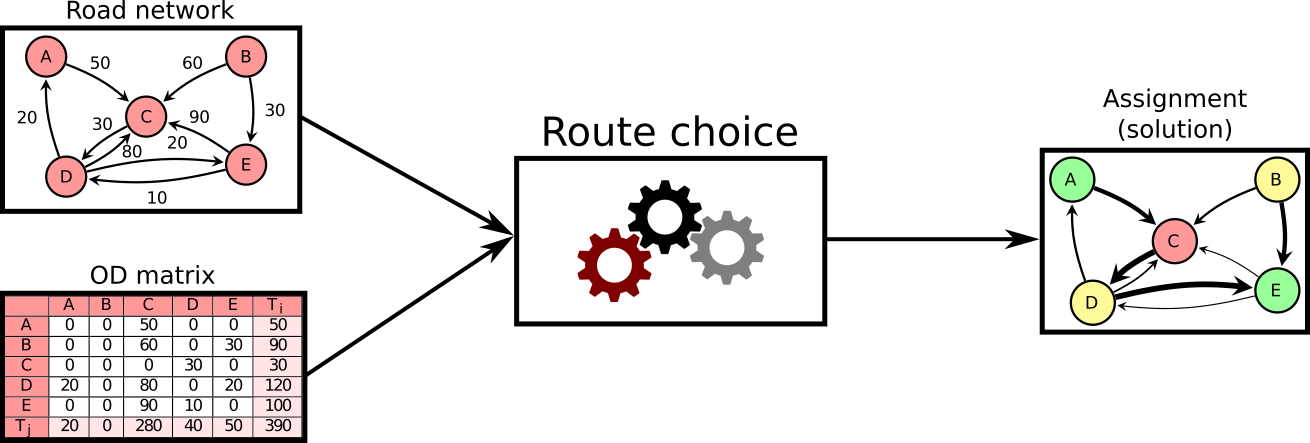}};
\node[anchor=center] (biasing) at (7.7,1)
    {\includegraphics[width=0.15\textwidth]{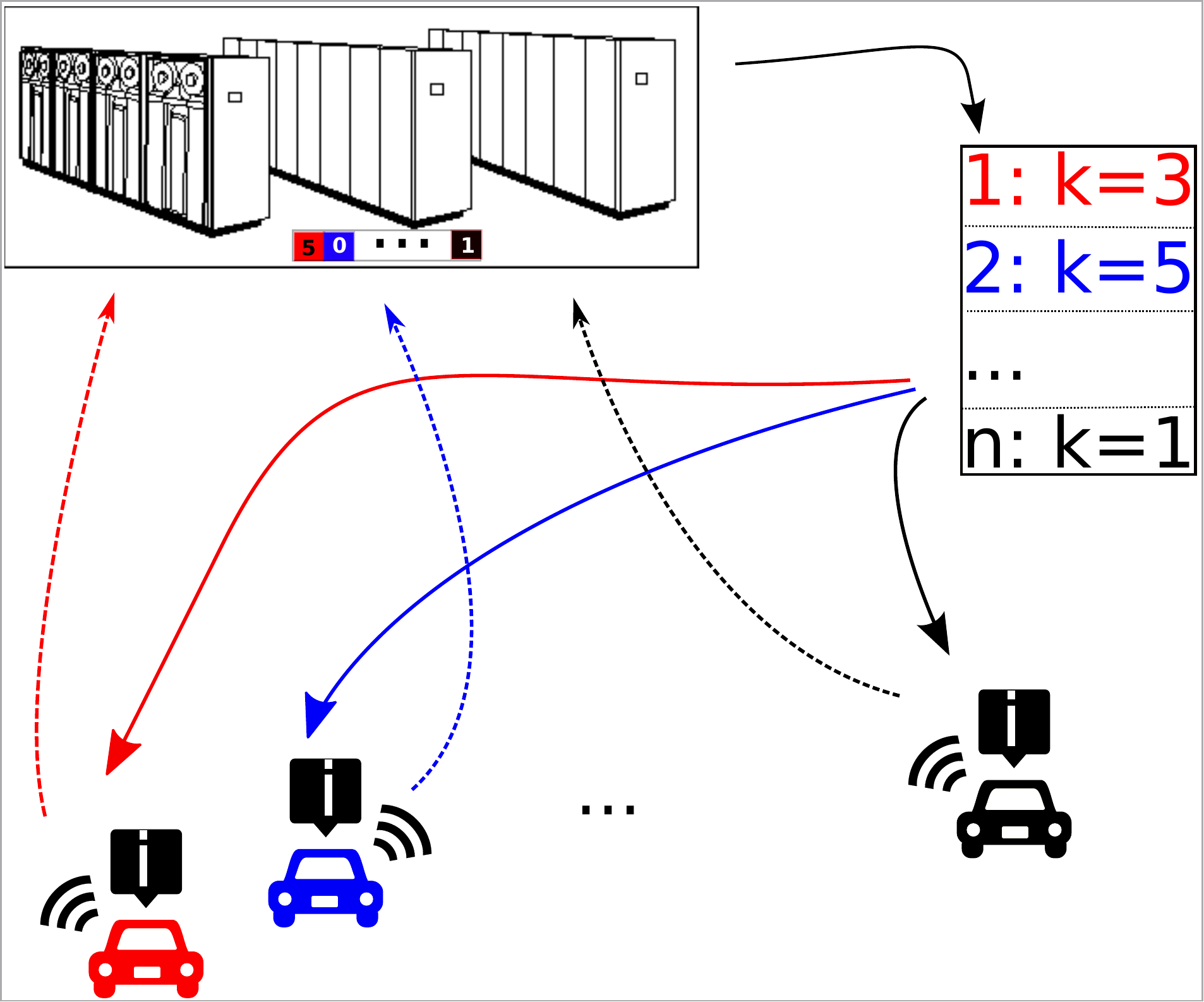}}; 
\node[anchor=center] (irc) at (11.5,1)
    {\includegraphics[width=.3\textwidth]{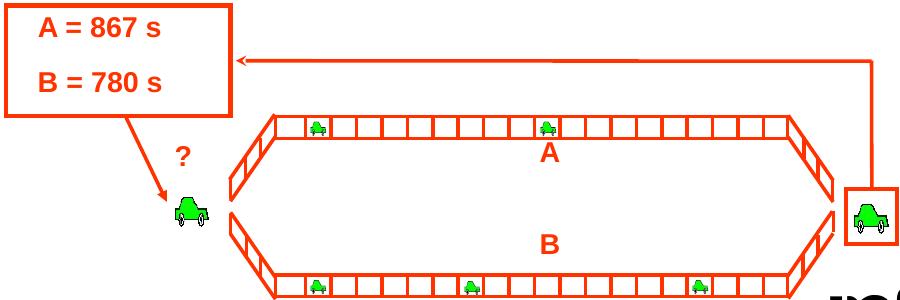}};
%     {\includegraphics[width=.17\textwidth]{/traffic/platoon3.jpeg}};
\node[anchor=center] (cell) at (9.4,0.4)
    {\includegraphics[width=.05\textwidth]{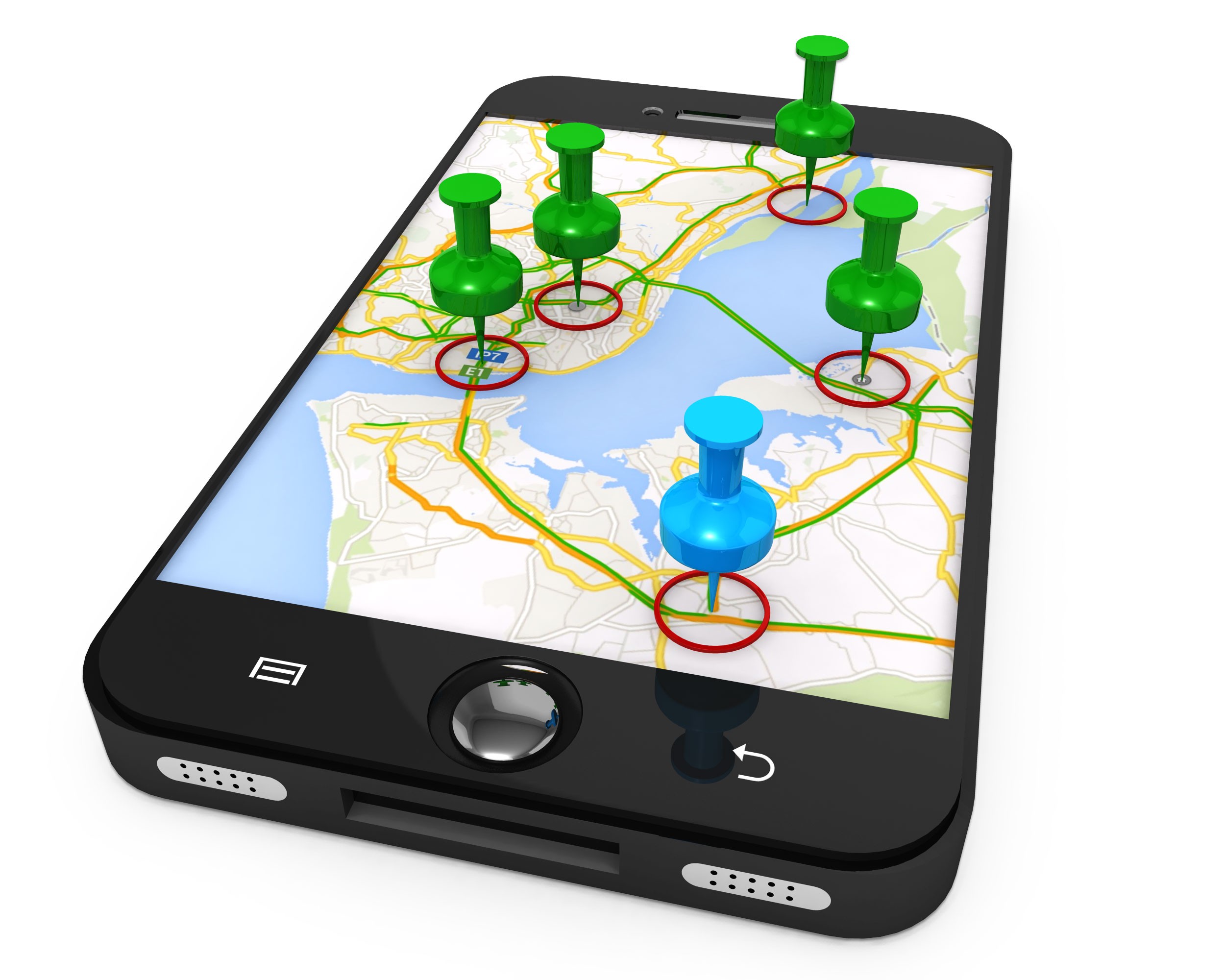}};
\end{tikzpicture}
}

While the previous section dealt with traffic control, this section
addresses the demand side, focusing on what seems to be a relevant issue for  all users of a road network, namely, how to get from A to B optimally
(less time, less cost, etc.). This problem has received several
approaches. In the case of the most recent research, the focus is on: dissemination of
information in a smart way, understanding the effect of such information, as well as 
behavioral changes by the drivers, how to learn to choose routes, 
how to disseminate information in order to guarantee a
certain level of system performance, and using  vehicular communication.

To achieve these objectives, the author has proposed several methods,
some pioneers in addressing the dissemination of information via
mobile devices when the smartphone  did not exist as we know 
today. Other methods  involved: game theory, 
\vx communication,  route choice via \ar, and effect of route recommendation on the 
alignment of the user and system optimum.

%***************************************************************************

\subsubsection{Human in the Loop: modeling route choice}
\label{sec:irc}

Section~\ref{sec:atis} discussed the  benefits of using an ATIS.
However, the dissemination of this type of technology brings with it the
need to consider the human being, when dealing with the loop between traffic control and allocation of resources. This question has received less attention, mainly due to computational issue. With the increase in the computation power of
processors, the advent of agent-based modeling, and also of
several multidisciplinary projects, several attempts to model the
problem of route choice were made, including those carried out by
author. In particular, agent-based modeling tries  to take into account the heterogeneity of such decisions; after all
each agent has its own particular way of making decisions. To do this, it was
necessary to develop  models that can represent the behavior of the
drivers, such as the use of  mental state models using the BDI  (Beliefs,
Desires, and Intentions) logic. Components of these
models can be: desires related to minimizing travel time, and beliefs
about the status and cost of each route or part of the road network used
by the agent. An application can be found in 
\cite{Bazzan+1999ki}. 

\subsubsection{Experimental Game Theory for Decision Making About Route Choice}

Understanding  human beings' reasoning on route choice is an open research question.
There are  no precise models for this. In order to investigate
this process, a useful tool is experimental game theory.
While classical game theory provides several tools for
modeling congestion games  (Section~\ref{sec:tap}), in experiments with human subjects
it is possible to observe whether and how they deviate from the results of
classical theory.

In a interdisciplinary project,  experiments were designed  where human subjects would iteratively
choose between two routes. After each iteration, they received 
information about corresponding outcomes. The main objective was to
study the effect of the dissemination of different types of information 
through mobile devices. It should be noted that this  idea
preceded the actual use of these devices - which would only come to the 
market some years later. The data collected in these experiments were the
basis for the formulation of heuristics for iterative route choice,
published in \cite{Kluegl&Bazzan2004jasss,Kluegl&Bazzan2004its}, where a simple form of \ar simulated the choices that were indeed made by the  human subjects. This work had important implications
 because, at that time, the typical means of disseminating
traffic  information was radio broadcast or
variable message panels (on the road). It needs to be stressed that there were no explicit recommendation of routes. More recently, other means of information dissemination exist, such as  Internet and geo-positioning, through services such as 
Waze and similars. These, in possession of the location of a 
significant number of  users of the service, recommend a route to the
user. One problem here is that the recommendation, when it is the same for
all users, can lead to an over use of the route when too many  users follow the recommendation. Simulations of this type of situation 
appear in \cite{Bazzan+2006lamas,Bazzan&Kluegl2005trc,Kluegl&Bazzan2004jasss}.

These are known issues in game theory. In problems related to minority games as in \cite{Bazzan&Kluegl2005trc}, it is known that in systems where each
participant tries to optimize his performance in an individual, myopic and greedy way, the 
overall performance is suboptimal. In the case of transportation networks,
assuming that vehicles try to avoid route A by opting for a route
B, the latter will have drastic performance loss, and there may be fluctuation and deterioration for all
participants, as in \cite{Wahle+2000}.
Specifically, this is a
question that underlies the so-called Braess paradox, originally
presented in \cite{Braess1968}, which represents a counter intuitive phenomenon: in a road network, when a new route connecting two points is built, it is possible that there is an increase in the
overall travel time, rather than a reduction. This happens because every decision by the 
drivers (based on their cost estimate) ignores the effects of
other drivers' decisions. That is, the
drivers, when trying to reduce their travel  times individually,
 end up increasing the  global travel time. 

As shown in  \cite{Bazzan&Kluegl2005trc}, the use of \ar makes drivers
adapt and learn how to distributed themselves among the available routes, thus improving the performance of the
system as a whole.

\subsubsection{Learning to Choose Routes}
\label{sec:rl4routing}

In the works mentioned in the previous section, \ar was used in a 
simplified way, without considering factors such as change of route during the trip, , agents' regret for inefficient choices,
simulation granularity (whether microscopic or macroscopic), and the search
for optimal system performance.

These points were addressed by the author with methods such as  \cite{Ramos&Grunitzki2015} (use of learning automata); in  \cite{Bazzan&Grunitzki2016} (modeling route choice as an stochastic game,
where the agent can change his route at each intersection);
in  
\cite{Ramos+2018trc} (choice route considering the past 
regret); and in
 \cite{Cagara+2015,Bazzan2019,Hasan+2016}.
In the case of these last three works, the objective was to align the optimum of the
system to the \ue. This problem, as mentioned
before and, in particular in the previous section, is linked to the paradox of
Braess, where adding resources to a system may degrade its performance.
To mitigate this problem,  \cite{Cagara+2015}
proposed the use of a distributed genetic algorithm. In the “islands” model, each vehicle can  communicate with others and  exchange solutions.
In the case of  \cite{Bazzan2019},  a genetic algorithm exchanges information with a
\ar process that runs at vehicle agents level. This synergy leads to more diverse solutions, which  not only accelerate the convergence of learning, but also lead to more efficient solutions at the global system level.
Aligning the  optimum of the
system to the \ue is achieved in \cite{Hasan+2016} by means of drivers sharing real-time information 
by means of an app that is used by members of a social network.

These works have shown that it is possible to tackle route choice from a decentralized point of view, by means of using \ar.
Our experience with such use has shown that the performance of a \ar method greatly varies according to the road network topology, demand distribution, as well as other factors.
Motivated by this, we have pursue the investigation of metrics that can characterize the traffic assignment problem according to the difficulty posed to the \ar.
These metrics were, firstly, based simply on centrality of nodes and edges of the network \cite{Batista&Bazzan2015,Galafassi&Bazzan2013}. Later, we proposed measures that compute the coupling between shortest routes \cite{Stefanello+2016,Oliveira+2017}.
Since none of these take into account the actual traffic assignment, a further measure was divised, based on the entropy of the distribution of routes, thus taking into account the actual demand in each route \cite{Redwan&Bazzan2020}.

\subsubsection{New Technologies for Guided Navigation}
\label{sec:c2x4nav}

A scheme that is becoming more popular is  congestion toll, i.e., charging a resource by its current  use as, for example, in Singapore where panels on some main roads
indicate the price to be payed for its use. In order to study the effects of this scheme, in
\cite{Tavares&Bazzan2012bwss} an approach was proposed that employs \ar in two classes of agents. The first one is the 
infrastructure through  road managers that learn how much to 
charge for each \vei. More congested roads should impose
  a price that  discourages its excessive use, so that traffic gets 
distributed over the network. On the other hand, driver agents also learn to
use routes that balance their cost and the agent's travel  time. In the  proposed model, the road network is treated as a \mas, where autonomous 
vehicles or vehicles plus their drivers have their local perception expanded through 
\vx communication. Road managers perceive the flow of vehicles on the roads they 
manage and try to maximize this flow through setting tolls. 
Further, in this work, a fraction of the
vehicle agents may act maliciously and try to reduce their travel times
disseminating false information to keep other drivers away from their
routes. 

The experiments were carried out in a microscopic simulator,
which replicates the behavior of drivers at various levels,
especially the one related to  route choice.
It was shown that a small fleet of malicious agents is
capable of  harming other drivers. In short,
this work anticipates scenarios that should occur in the near future.

Three other, more recent, works by the author also address
communication, either among vehicles or between vehicles and  infrastructure.

In \cite{Huang+2020}, connected vehicles can communicate in order to coordinate
their route choices through an  app. This  app, having the 
information collected from vehicles that subscribe  this service,
recommends routes that leads the system close to its optimum, rather than to the \ue.

Both \cite{Santos&Bazzan2021} and \cite{Bazzan&Kluegl2020} aim to investigate how to speed up the process of
agent learning in terms of choosing routes using
\vx communication. In the former, vehicles inform the
road infrastructure (link managers) how much time it took to cross a link or a portion of the network. Managers aggregate such information and distribute it to
vehicles that are yet to decide which link to use to reach their
destinations. In the latter work, vehicles inform their travel times to an app that shares the rewards (travel times) obtained when a particular
route was used. 

In both works, it was shown that the learning task is achieved  quicker through communication. Note that, contrary to previous works such as  \cite{Wahle+2000},
these two most recent works are about sharing just local information.
This way, such  scheme avoids the aforementioned phenomenon where sharing
the same information for all agents can lead to drop in performance.

\subsubsection{Guided Navigation: main results}

\begin{minipage}[t]{0.95\linewidth} 
\begin{tcolorbox}[colback=info,colframe=blue] %,title=Principal resultado]
Development and testing of various methods that model agents \
individually in order to investigate the effects of  disseminating
information for route choice, as well as aligning user and global optima. This is done through different schemes,
including  congestion charge and synergy
between machine learning techniques. 
\end{tcolorbox} 
\end{minipage}

\subsection{Integrating Intelligent Control and Guided Navigation}
\label{sec:coadap}

\noindent
\resizebox{\textwidth}{!}{
\begin{tikzpicture}
\draw (-1,0) [fill=coadap,draw=white] rectangle (14,2);
\node[anchor=east] (marl) at (13,1)
	{\begin{tikzpicture} 
		\draw[fill=gray!30,draw=white] (10,4) -- (10.5,4) -- (11,2.5) -- (10.5,2.5)  -- cycle;
		\draw[color=black, style=dashed] (10.25,4) -- (10.75,2.5); 
		\node at (11,4) {\includegraphics[width=.03\textwidth]{figs/signal}};
		\node at (11.2,3.8) {\includegraphics[width=.04\textwidth]{figs/signal}};
		\node at (11.5,3.5) {\includegraphics[width=.06\textwidth]{figs/signal}};
		\node at (12.2,3.1) {\includegraphics[width=.1\textwidth]{figs/signal}};
		\node (s) at (13,3.5) {\scriptsize state?};
		\node at (13,3.0) {\scriptsize action: $a$};
		\node at (13,2.5) {\scriptsize reward: $r$};
		\draw[<->] (bigger.north) to [out=30,in=90]   (s.west);
	\end{tikzpicture}};
\node[anchor=center] (irc) at (1.5,1)
    {\includegraphics[width=.3\textwidth]{/traffic/irc.jpeg}};
\draw[<->,thick,double] (marl) -- (irc) node at (7.2,0.7){co-learning} ;  % futuro: 2 bend arrows
\end{tikzpicture}
}

Sections \ref{sec:control} and \ref{sec:routing} presented several
methods proposed and developed by the author and colleagues to address issues of
intelligent control of traffic signals and route choice respectively.
However, in the real world, not only these two tasks do  occur
simultaneously but also are highly coupled. Clearly, the
learning task carried out at traffic signal controllers affects the learning task  of
drivers and vice versa. Therefore, it is important to consider the
simultaneous adaptation of the two classes of agents, i.e., co-learning.

In 
\cite{Bazzan+2008alamas} the theoretical basis for using 
\ar in both classes of agents was proposed. However, the underlying simulation model used there was not fully
microscopic. 
Thus, in  \cite{Lemos&Bazzan2019,Lemos+2018} the original proposal has been extended to include a microscopic simulation environment, which led to some challenges. First, the
problem of having two classes of agents representing the supply and the
demand adapting simultaneously makes the problem more complex
computationally speaking, since several convergence guarantees are lost. 
Second, the learning tasks become
more complex because the actions of the agents are highly coupled. 
A further challenge is the fact that the natures of these learning tasks are 
different; for instance, a driver’s goal is to minimize his
individual travel time, while the goal of a traffic signal controller is to reduce queues
locally.

Therefore, in  \cite{Lemos&Bazzan2019,Lemos+2018} an approach was proposed where the drivers' learning task is  based on repeated games; for the traffic signals, the learning task is based on stochastic games. 
While the former is based on episodes that are not synchronous, 
traffic signals have an infinite learning horizon, i.e., there are no proper episodes. 
Finally, another challenge regards the  microscopic simulation. The
experiments were carried out using a microscopic simulator and a
grid network with 32 traffic signals. It was shown that co-learning reduced travel times and queues of 
vehicles at the intersections.

\vspace{0.5cm}
\begin{minipage}[t]{0.95\linewidth} 
\begin{tcolorbox}[colback=coadap,colframe=PineGreen,title=Main result]
Development of a non-trivial mechanism that integrates \ar being performed in two classes of agents, with distinct  characteristics and objectives.
\end{tcolorbox} 
\end{minipage}

\subsection{Ongoing work}
\label{sec:ongoing}

As mentioned in Section~\ref{sec:open},  in the real world, it is rarely the case that one deals with only one objective to be optimized (such as travel time or queue length at an intersection).
Typically, an optimization processes involve several variables.
Therefore, there is  a clear gap in the literature.
The work by \cite{Reymond&Nowe2019} 
uses a traffic signal control (single intersection)
scenario to illustrate the use of a multiobjective approach, where   objectives are to maximize
the traffic flow and minimize  waiting times.
However, the paper does not give all the details, since such scenario is used together with others as illustration.

In short, it seems that there are just few  works  capable of dealing with multiobjective
\ar in urban mobility scenarios.
In the case of vehicles, the existing methods
only address drivers who seek to minimize their travel times,
disregarding other factors such as toll, emissions, or battery consumption.
In the case of traffic signal controllers, it is clear that there are several measures that could be optimized simultaneously, even if some of them are correlated.

%%%%%%%% fig do COMA
\begin{figure*}[t]
    \centering
	\includegraphics[width=0.9\textwidth]{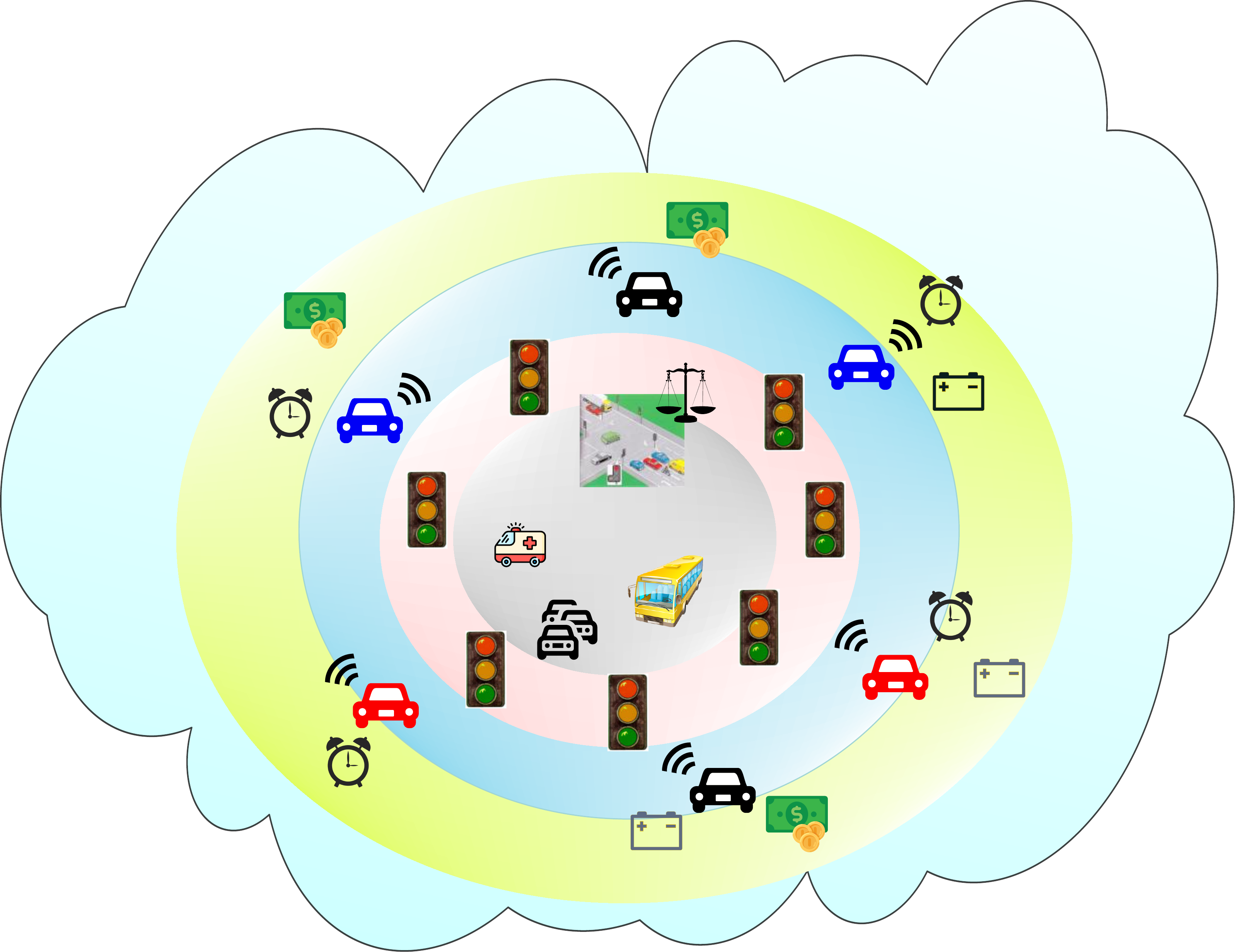}
	\caption{Communication as a Substrate for Machine Learning and Making
Multiples Decision
Goals.}
	\label{fig:MOcircles}
\end{figure*}

This is where 
our ongoing work is currently focusing.
Figure~\ref{fig:MOcircles} illustrates this vision with a scheme, in which most of the elements have the ability to communicate, 
either through  \vx communication, or through the cloud, or another
protocol. 

With regard to traffic signal controllers (red circle in the figure),
in addition to managing traffic at local level, they must: (i) communicate
in order to interact cooperatively, and (ii) make decisions that
include multiple, sometimes conflicting objectives. The figure shows
some of these objectives (central circle, gray): minimize queues,
reduce waiting times at the intersection, reduce travel times (not only locally but throughout the entire trip), give priority to emergency vehicles  and public transportation, and allocate green times
among the phases in a fair way. 

In the case of vehicles (blue circle in the figure),
these being equipped with communication devices  (all of them or only partially), it is necessary to formulate communication  protocols that are effective and efficient. 

Finally, as
already mentioned, there is a gap with respect to \ar when 
agents must make decisions based on multiple objectives. In the case of
vehicles, in the real world, their drivers aim not only at minimizing their
travel times (as commonly assumed), but also other goals
(green circle in the figure): minimize costs involved with  tolls,
increase battery life (if electric vehicles) by selecting
routes that can regenerate battery capacity, etc. 
Further, it is necessary to consider 
heterogeneous agents with respect to their preferences and preferences, which is an
advantage of an agent-based modeling.

In a work in progress, algorithms for multiobjective
\ar are being proposed, which address part of the aforementioned agenda.
Such algorithms are not trivial since they do not just formulate the various objectives as a (linear or
non-linear) combination of the objectives. This is important in order to  guarantee that 
all solutions at the Pareto front will be found. 

Specifically, classical algorithms for single-agent, single-objective
\ar such as Q-learning \cite{Watkins&Dayan1992} and UCB \cite{Auer+2002ucb} are
being extended and tested in route choice scenarios, where the
objectives are not only to minimize travel time, but also the toll expenditure.
Preliminary results show a good performance
of the new algorithms, as they achieve the same result as centralized algorithms
such as \cite{Raith+2014}), while being faster.

\vspace{0.5cm}
\begin{minipage}[t]{0.95\linewidth} 
\begin{tcolorbox}[colback=Tan,colframe=brown,title=Main result]
Extension of \ar algorithms for  multi-objective scenarios, especially for route choice.
\end{tcolorbox} 
\end{minipage}

\section{Conclusion}

The agenda around improving urban mobility is a
priority of municipal authorities. In this paper, some
among the various aspects of this agenda were discussed. It has been shown that it is possible to 
improve the overall efficiency of the transportation system through  artificial intelligence methods and
technologies. This refers both to traffic 
control measures, as well as measures related to the dissemination of
information to drivers and also to connected and autonomous vehicles. 

Specifically, this work has contributed and continues to contribute to the
development of new techniques that were described here. The highlights are:

\begin{enumerate}
	\item New techniques have been developed involving \marl. This allows agents to adapt
    in a decentralized way, that is, without central control. Recall that centralization can be costly both computationally and in terms of
    infrastructure, since it may require cabling and/or communication with
    a central entity. The fact that learning involves dozens or even
    even thousands of agents (if traffic signals or drivers are involved,
    respectively), makes the problem more complex due to the non-stationary nature of the learning tasks. To mitigate these problems, the author developed
    hierarchical methods or synergies with other methods have led to a
    significant increase in the speed of learning, something that is a
    critical factor in dynamic environments, where agents must
    adapt as quick as possible.
    
	\item Methods have been developed and tested that simulate and anticipate
    effects of policies for the dissemination of information to travelers.
    These methods showed the feasibility of obtaining better performance,
    in terms of the system as a whole.
	
	\item  Methods based on \ar were able to reduce queues (in the case of 
    traffic signal controllers) and travel times (vehicles).
%	\item Software for microscopic traffic simulation was built,    in open source.
\end{enumerate}

\section*{Acknowledgments}
Ana Bazzan is partially supported by CNPq (grant 307215/2017-2).
This work was partially supported by   CAPES (Coordena\c c\~ao de Aperfei\c coamento de Pessoal de N\'ivel Superior - Brasil) - Finance Code 001.

\bibliographystyle{unsrt}
\bibliography{stringDefs,AL,MZ,temp,OURS}

\begin{thebibliography}{10}

\bibitem{Mitchell+2010}
William~J. Mitchell, Christopher~E. Borroni-Bird, and Lawrence~D. Burns.
\newblock {\em Reinventing the Automobile}.
\newblock MIT Press, Cambridge, MA, 2010.

\bibitem{Wahle+2000}
J.~Wahle, Ana L.~C. Bazzan, F.~Kl\"ugl, and M.~Schreckenberg.
\newblock Decision dynamics in a traffic scenario.
\newblock {\em Physica A}, 287({3--4}):669--681, 2000.

\bibitem{Wahle+2002}
J.~Wahle, Ana L.~C. Bazzan, and F.~Kl\"ugl.
\newblock The impact of real time information in a two route scenario using
  agent based simulation.
\newblock {\em Transportation Research Part C: Emerging Technologies},
  10(5--6):73--91, 2002.

\bibitem{Wachs2010}
Martin Wachs.
\newblock Transportation policy, poverty, and sustainability: History and
  future.
\newblock {\em Transportation Research Record}, 2163(1):5--12, 2010.

\bibitem{Bazzan&Kluegl2007}
Ana L.~C. Bazzan and Franziska Kl\"ugl.
\newblock Sistemas inteligentes de transporte e tr\'afego: uma abordagem de
  tecnologia da informa\c c\~ao.
\newblock In Tomasz Kowaltowski and Karin~Koogan Breitman, editors, {\em Anais
  das Jornadas de Atualiza\c c\~ao em Inform\'atica}, chapter~8. {SBC}, July
  2007.

\bibitem{Bazzan&Kluegl2013book}
Ana L.~C. Bazzan and Franziska Kl\"ugl.
\newblock {\em Introduction to Intelligent Systems in Traffic and
  Transportation}, volume~7 of {\em Synthesis Lectures on Artificial
  Intelligence and Machine Learning}.
\newblock Morgan and Claypool, 2013.

\bibitem{Bazzan&Kluegl2014}
Ana L.~C. Bazzan and Franziska Kl\"ugl.
\newblock A review on agent-based technology for traffic and transportation.
\newblock {\em The Knowledge Engineering Review}, 29(3):375--403, 2014.

\bibitem{Kluegl&Bazzan2012}
Franziska Kl\"ugl and Ana L.~C. Bazzan.
\newblock Agent-based modeling and simulation.
\newblock {\em AI Magazine}, 33(3):29--40, 2012.

\bibitem{Wardrop1952}
John~Glen Wardrop.
\newblock Some theoretical aspects of road traffic research.
\newblock {\em Proceedings of the Institution of Civil Engineers, Part II},
  1(36):325--362, 1952.

\bibitem{Roughgarden&Tardos2002}
Tim Roughgarden and {\'E}va Tardos.
\newblock How bad is selfish routing?
\newblock {\em J. ACM}, 49(2):236--259, 2002.

\bibitem{Wooldridge2009}
M.~J. Wooldridge.
\newblock {\em An Introduction to MultiAgent Systems}.
\newblock John Wiley \& Sons, Chichester, 2009.
\newblock Second edition.

\bibitem{Bazzan2010}
Ana L.~C. Bazzan.
\newblock {IA} multiagente: Mais intelig\^encia, mais desafios.
\newblock In Wagner Meira, Jr. and Andr\'e C. P. L.~F. de~Carvalho, editors,
  {\em Atualiza\c c\~oes em inform\'atica 2010}, chapter~3, pages 111--159.
  PUC-Rio, Rio de Janeiro, July 2010.

\bibitem{Robertson1969}
D.~I. Robertson.
\newblock {TRANSYT}: A traffic network study tool.
\newblock Rep. LR 253, Road Res. Lab., London, 1969.

\bibitem{Transyt1988}
TRANSYT-7F.
\newblock {\em TRANSYT-7F User's Manual}.
\newblock Transportation Research Center, University of Florida, 1988.

\bibitem{Hunt+1981}
P.~B. Hunt, D.~I. Robertson, R.~D. Bretherton, and R.~I. Winton.
\newblock {SCOOT} - a traffic responsive method of coordinating signals.
\newblock {TRRL} Lab. Report 1014, Transport and Road Research Laboratory,
  Berkshire, 1981.

\bibitem{Robertson&Bretherton1991}
Dennis~I. Robertson and R.~David Bretherton.
\newblock Optimizing networks of traffic signals in real time - the {SCOOT}
  method.
\newblock {\em IEEE Transactions on Vehicular Technology}, 40(1):11--15,
  February 1991.

\bibitem{Lowrie1982}
P.~Lowrie.
\newblock The {S}ydney coordinate adaptive traffic system - principles,
  methodology, algorithms.
\newblock In {\em Proceedings of the International Conference on Road Traffic
  Signalling}, Sydney, Australia, 1982.

\bibitem{Diakaki+2003}
C.~Diakaki, V.~Dinopoulou, K.~Aboudolas, M.~Papageorgiou, E.~Ben-Shabat,
  E.~Seider, and A.~Leibov.
\newblock Extensions and new applications of the traffic signal control
  strategy {TUC}.
\newblock In {\em Proc. of the 82nd Annual Meeting of the Transportation
  Research Board}, pages 12--16, January 2003.

\bibitem{Bazzan2009}
Ana L.~C. Bazzan.
\newblock Opportunities for multiagent systems and multiagent reinforcement
  learning in traffic control.
\newblock {\em Autonomous Agents and Multiagent Systems}, 18(3):342--375, June
  2009.

\bibitem{Mannion+2016}
Patrick Mannion, Jim Duggan, and Enda Howley.
\newblock An experimental review of reinforcement learning algorithms for
  adaptive traffic signal control.
\newblock In Thomas Leo~McCluskey, Apostolos Kotsialos, P.~J\"org M\"uller,
  Franziska Kl\"ugl, Omer Rana, and Ren\'e Schumann, editors, {\em Autonomic
  Road Transport Support Systems}, pages 47--66. Springer International
  Publishing, Cham, May 2016.

\bibitem{Wei+2021}
Hua Wei, Guanjie Zheng, Vikash Gayah, and Zhenhui Li.
\newblock Recent advances in reinforcement learning for traffic signal control:
  A survey of models and evaluation.
\newblock {\em SIGKDD Explor. Newsl.}, 22(2):12--18, 2021.

\bibitem{Yau+2017}
Kok-Lim~Alvin Yau, Junaid Qadir, Hooi~Ling Khoo, Mee~Hong Ling, and Peter
  Komisarczuk.
\newblock A survey on reinforcement learning models and algorithms for traffic
  signal control.
\newblock {\em ACM Comput. Surv.}, 50(3), 2017.

\bibitem{Wiering2000}
Marco Wiering.
\newblock Multi-agent reinforcement learning for traffic light control.
\newblock In {\em Proceedings of the Seventeenth International Conference on
  Machine Learning ({ICML} 2000)}, pages 1151--1158, 2000.

\bibitem{Taale+2015}
Henk Taale, Joost {van Kampen}, and Serge Hoogendoorn.
\newblock Integrated signal control and route guidance based on back-pressure
  principles.
\newblock {\em Transportation Research Procedia}, 10:226--235, 2015.

\bibitem{Adler&Blue1998}
J.~L. Adler and V.~J. Blue.
\newblock Toward the design of intelligent traveller information systems.
\newblock {\em Transportation Research Part C}, 6:157--172, 1998.

\bibitem{Ben-Akiva+1991}
M.~Ben-Akiva, A.~de~Palma, and I.~Kaysi.
\newblock Dynamic network models and driver information systems.
\newblock {\em Transp. Res. A}, 25(5):251--266, 1991.

\bibitem{Bonsall1992}
P.~W. Bonsall.
\newblock The influence of route guidance advice on route choice in urban
  networks.
\newblock {\em Transportation}, 19(1), 1992.

\bibitem{Mahmassani&Chen1991}
H.~S. Mahmassani and P.~S. Chen.
\newblock Comparative assessment of origin-based and en route real-time
  information under alternative user behavior rules.
\newblock {\em Transportation Research Record}, 1306:69--81, 1991.

\bibitem{Sharon+2017}
Guni Sharon, Josiah~P Hanna, Tarun Rambha, Michael~W Levin, Michael Albert,
  Stephen~D Boyles, and Peter Stone.
\newblock Real-time adaptive tolling scheme for optimized social welfare in
  traffic networks.
\newblock In S.~Das, E.~Durfee, K.~Larson, and M.~Winikoff, editors, {\em Proc.
  of the 16th International Conference on Autonomous Agents and Multiagent
  Systems (AAMAS 2017)}, pages 828--836, S\~ao Paulo, May 2017. IFAAMAS.

\bibitem{Buriol+2010}
Luciana~S. Buriol, Michael~J. Hirsh, Panos~M. Pardalos, Tania Querido,
  Mauricio~G.C. Resende, and Marcus Ritt.
\newblock A biased random-key genetic algorithm for road congestion
  minimization.
\newblock {\em Optimization Letters}, 4:619--633, 2010.

\bibitem{Arnott+1990}
R.~Arnott, A.~de~Palma, and R.~Lindsey.
\newblock Departure time and route choice for the morning commute.
\newblock {\em Transportation Research, Part B}, 24:209--228, 1990.

\bibitem{Kobayashi&Do2005}
Kiyoshi Kobayashi and Myungsik Do.
\newblock The informational impacts of congestion tolls upon route traffic
  demands.
\newblock {\em Transportation Research, Part A}, 39(7--9):651--670,
  August-November 2005.

\bibitem{Dia&Panwai2014}
H.~Dia and S.~Panwai.
\newblock {\em Intelligent Transport Systems: Neural Agent (Neugent) Models of
  Driver Behaviour}.
\newblock LAP Lambert Academic Publishing, 2014.

\bibitem{Barthelemy&Carletti2017}
Johan Barth{\'e}lemy and Timoteo Carletti.
\newblock A dynamic behavioural traffic assignment model with strategic agents.
\newblock {\em Transportation Research Part C: Emerging Technologies},
  85:23--46, 2017.

\bibitem{Dias+2014}
Jos\'e~Capela Dias, Penousal Machado, Daniel~Castro Silva, and Pedro~Henriques
  Abreu.
\newblock An inverted ant colony optimization approach to traffic.
\newblock {\em Engineering Applications of Artificial Intelligence},
  36(0):122--133, July 2014.

\bibitem{Claes+2011}
Rutger Claes, Tom Holvoet, and Danny Weyns.
\newblock A decentralized approach for anticipatory vehicle routing using
  delegate multiagent systems.
\newblock {\em IEEE Transactions on Intelligent Transportation Systems},
  12(2):364--373, March 2011.

\bibitem{Guestrin+2002}
Carlos Guestrin, Michail~G. Lagoudakis, and Ronald Parr.
\newblock Coordinated reinforcement learning.
\newblock In Claude Sammut and Achim~G. Hoffmann, editors, {\em Proceedings of
  the Nineteenth International Conference on Machine Learning (ICML)}, pages
  227--234, San Francisco, CA, USA, 2002. Morgan Kaufmann.

\bibitem{Hu&Wellman1998}
Junling Hu and Michael~P. Wellman.
\newblock Multiagent reinforcement learning: Theoretical framework and an
  algorithm.
\newblock In {\em Proc. 15th International Conf. on Machine Learning}, pages
  242--250. Morgan Kaufmann, 1998.

\bibitem{Melo&Ribeiro2010}
Francisco Melo and M.~Ribeiro.
\newblock Coordinated learning in multiagent {MDPs} with infinite state-space.
\newblock {\em Autonomous Agents and Multi-Agent Systems}, 21(3):321--367,
  2010.

\bibitem{Stone2007}
Peter Stone.
\newblock Multiagent learning is not the answer. {I}t is the question.
\newblock {\em Artificial Intelligence}, 171(7):402--405, May 2007.

\bibitem{Zhang+2008}
Chongjie Zhang, Sherief Abdallah, and Victor~R. Lesser.
\newblock Efficient multi-agent reinforcement learning through automated
  supervision (extended abstract).
\newblock In Lin Padgham, David Parkes, J.~M\"uller, and Simon Parsons,
  editors, {\em Proceedings of the 7th International Joint Conference on
  Autonomous Agents and Multiagent Systems}, volume~3, pages 1365--1368,
  Estoril, 2008.

\bibitem{Radulescu+2020}
Roxana R\v{a}dulescu, Patrick Mannion, Diederik Roijers, and Ann Now\'e.
\newblock Multi-objective multi-agent decision making: a utility-based analysis
  and survey.
\newblock {\em Autonomous Agents and Multi-Agent Systems}, 34, 04 2020.

\bibitem{Hayes+2021}
Conor~F. Hayes, Roxana Radulescu, Eugenio Bargiacchi, Johan
  K{\"{a}}llstr{\"{o}}m, Matthew Macfarlane, Mathieu Reymond, Timothy
  Verstraeten, Luisa~M. Zintgraf, Richard Dazeley, Fredrik Heintz, Enda Howley,
  Athirai~A. Irissappane, Patrick Mannion, Ann Now{\'{e}}, Gabriel
  de~Oliveira~Ramos, Marcello Restelli, Peter Vamplew, and Diederik~M. Roijers.
\newblock A practical guide to multi-objective reinforcement learning and
  planning.
\newblock {\em CoRR}, abs/2103.09568, 2021.

\bibitem{Lopez+2018}
Pablo~Alvarez Lopez, Michael Behrisch, Laura Bieker-Walz, Jakob Erdmann,
  Yun-Pang Fl{\"o}tter{\"o}d, Robert Hilbrich, Leonhard L{\"u}cken, Johannes
  Rummel, Peter Wagner, and Evamarie Wie{\ss}ner.
\newblock Microscopic traffic simulation using sumo.
\newblock In {\em The 21st IEEE International Conference on Intelligent
  Transportation Systems}, 2018.

\bibitem{Nagel&Schreckenberg1992}
K.~Nagel and M.~Schreckenberg.
\newblock A cellular automaton model for freeway traffic.
\newblock {\em Journal de Physique I}, 2:2221, 1992.

\bibitem{Bazzan+2010atss}
Ana L.~C. Bazzan, Maicon de Brito~{\relax do} Amarante, Tiago Sommer, and
  Alexander~J. Benavides.
\newblock {ITSUMO}: an agent-based simulator for {ITS} applications.
\newblock In Rosaldo Rossetti, Henry Liu, and Shuming Tang, editors, {\em Proc.
  of the 4th Workshop on Artificial Transportation Systems and Simulation}.
  IEEE, September 2010.

\bibitem{Silva+2005}
Bruno Castro~{\relax da} Silva, Ana L.~C. Bazzan, Gustavo.~K. Andriotti, and
  Denise~de Oliveira.
\newblock {ITSUMO:} an intelligent transportation system for urban mobility.
\newblock In {\em Proceedings of the Optimization of Urban Traffic Systems},
  Lecture Notes in Computer Science, pages 224--235, Guadalajara, Mexico, 2005.
  Springer-Verlag.

\bibitem{Bazzan2005}
Ana L.~C. Bazzan.
\newblock A distributed approach for coordination of traffic signal agents.
\newblock {\em Autonomous Agents and Multiagent Systems}, 10(1):131--164, March
  2005.

\bibitem{Oliveira+2004}
Denise de~Oliveira, Paulo~R. Ferreira, Jr., Ana L.~C. Bazzan, and Franziska
  Kl\"ugl.
\newblock A swarm-based approach for selection of signal plans in urban
  scenarios.
\newblock In {\em Proceedings of Fourth International Workshop on Ant Colony
  Optimization and Swarm Intelligence - {ANTS} 2004}, volume 3172 of {\em
  Lecture Notes in Computer Science}, pages 416--417, Berlin, Germany, 2004.

\bibitem{Oliveira+2005}
Denise~de Oliveira, Ana L.~C. Bazzan, and V.~Lesser.
\newblock Using cooperative mediation to coordinate traffic lights: a case
  study.
\newblock In Frank Dignum, Virginia Dignum, Sven Koenig, Sarit Kraus,
  Munindar~P. Singh, and Michael Wooldridge, editors, {\em Proceedings of the
  4th International Joint Conference on Autonomous Agents and Multi Agent
  Systems {(AAMAS)}}, pages 463--470. New York, {IEEE} Computer Society, July
  2005.

\bibitem{Silva+2006icml}
Bruno Castro~da Silva, Eduardo~W. Basso, Ana L.~C. Bazzan, and Paulo~M. Engel.
\newblock Dealing with non-stationary environments using context detection.
\newblock In William~W. Cohen and Andrew Moore, editors, {\em Proceedings of
  the 23rd International Conference on Machine Learning {ICML}}, pages
  217--224. New York, {ACM} Press, June 2006.

\bibitem{Bazzan+2010eaai}
Ana L.~C. Bazzan, Denise de~Oliveira, and Bruno~C. da~Silva.
\newblock Learning in groups of traffic signals.
\newblock {\em Eng. Applications of Art. Intelligence}, 23:560--568, 2010.

\bibitem{Abdoos+2013eaai}
Monireh Abdoos, Nasser Mozayani, and Ana~L.C. Bazzan.
\newblock Holonic multi-agent system for traffic signals control.
\newblock {\em Engineering Applications of Artificial Intelligence},
  26(5--6):1575--1587, 2013.

\bibitem{Watkins&Dayan1992}
Christopher J. C.~H. Watkins and Peter Dayan.
\newblock {Q}-learning.
\newblock {\em Machine Learning}, 8(3):279--292, 1992.

\bibitem{Abdoos&Bazzan2021}
Monireh Abdoos and Ana~L.C. Bazzan.
\newblock Hierarchical traffic signal optimization using reinforcement learning
  and traffic prediction with long-short term memory.
\newblock {\em Expert Systems with Applications}, page 114580, 2021.

\bibitem{Bazzan+1999ki}
Ana L.~C. Bazzan, J.~Wahle, and F.~Kl\"ugl.
\newblock Agents in traffic modelling - from reactive to social behavior.
\newblock In {\em Advances in Artificial Intelligence}, number 1701 in Lecture
  Notes in Artificial Intelligence, pages 303--306, Berlin/Heidelberg, 1999.
  Springer.
\newblock Extended version appeared in Proc. of the {U.K.} Special Interest
  Group on Multi-Agent Systems {(UKMAS),} Bristol, {UK.}

\bibitem{Kluegl&Bazzan2004jasss}
F.~Kl\"ugl and Ana L.~C. Bazzan.
\newblock Route decision behaviour in a commuting scenario.
\newblock {\em Journal of Artificial Societies and Social Simulation}, 7(1),
  2004.

\bibitem{Kluegl&Bazzan2004its}
F.~Kl\"ugl and Ana L.~C. Bazzan.
\newblock Simulation studies on adaptative route decision and the influence of
  information on commuter scenarios.
\newblock {\em Journal of Intelligent Transportation Systems: Technology,
  Planning, and Operations}, 8(4):223--232, October/December 2004.

\bibitem{Bazzan+2006lamas}
Ana L.~C. Bazzan, M.~Fehler, and F.~Kl\"ugl.
\newblock Learning to coordinate in a network of social drivers: The role of
  information.
\newblock In Karl Tuyls, Pieter~{Jan't} Hoen, Katja Verbeeck, and Sandip Sen,
  editors, {\em Proceedings of the International Workshop on Learning and
  Adaptation in {MAS} {(LAMAS 2005)}}, number 3898 in Lecture Notes in
  Artificial Intelligence, pages 115--128, 2006.

\bibitem{Bazzan&Kluegl2005trc}
Ana L.~C. Bazzan and Franziska Kl\"ugl.
\newblock Case studies on the {B}raess paradox: simulating route recommendation
  and learning in abstract and microscopic models.
\newblock {\em Transportation Research C}, 13(4):299---319, August 2005.

\bibitem{Braess1968}
D.~Braess.
\newblock {\"U}ber ein {P}aradoxon aus der {V}erkehrsplanung.
\newblock {\em Unternehmensforschung}, 12:258, 1968.

\bibitem{Ramos&Grunitzki2015}
Gabriel {\relax de}~O. Ramos and Ricardo Grunitzki.
\newblock An improved learning automata approach for the route choice problem.
\newblock In Fernando Koch, Felipe Meneguzzi, and Kiran Lakkaraju, editors,
  {\em Agent Technology for Intelligent Mobile Services and Smart Societies},
  volume 498 of {\em Communications in Computer and Information Science}, pages
  56--67. Springer Berlin Heidelberg, 2015.

\bibitem{Bazzan&Grunitzki2016}
Ana L.~C. Bazzan and R.~Grunitzki.
\newblock A multiagent reinforcement learning approach to en-route trip
  building.
\newblock In {\em 2016 International Joint Conference on Neural Networks
  (IJCNN)}, pages 5288--5295, July 2016.

\bibitem{Ramos+2018trc}
Gabriel {\relax de}~O. Ramos, Ana L.~C. Bazzan, and Bruno~C. {\relax da}~Silva.
\newblock Analysing the impact of travel information for minimising the regret
  of route choice.
\newblock {\em Transportation Research Part C: Emerging Technologies},
  88:257--271, Mar 2018.

\bibitem{Cagara+2015}
Daniel Cagara, Bj\"orn Scheuermann, and Ana~L.C. Bazzan.
\newblock Traffic optimization on islands.
\newblock In {\em 7th IEEE Vehicular Networking Conference (VNC 2015)}, pages
  175--182, Kyoto, Japan, December 2015. IEEE.

\bibitem{Bazzan2019}
Ana L.~C. Bazzan.
\newblock Aligning individual and collective welfare in complex socio-technical
  systems by combining metaheuristics and reinforcement learning.
\newblock {\em Eng. Appl. of {AI}}, 79:23--33, 2019.

\bibitem{Hasan+2016}
M.~R. Hasan, A.~L.~C. Bazzan, E.~Friedman, and A.~Raja.
\newblock A multiagent solution to overcome selfish routing in transportation
  networks.
\newblock In {\em 2016 IEEE 19th International Conference on Intelligent
  Transportation Systems (ITSC)}, pages 1850--1855, Nov 2016.

\bibitem{Batista&Bazzan2015}
Rodrigo {\relax de}~Abreu Batista and Ana L.~C. Bazzan.
\newblock Identification of central points in road networks using betweenness
  centrality combined with traffic demand.
\newblock {\em Polibits}, 52:85--91, 2015.

\bibitem{Galafassi&Bazzan2013}
Cristiano Galafassi and Ana L.~C. Bazzan.
\newblock Analysis of traffic behavior in regular grid and real world networks.
\newblock In {\em The Fifth International Workshop on Emergent Intelligence on
  Networked Agents (WEIN)}, 2013.

\bibitem{Stefanello+2016}
Fernando Stefanello, Bruno~C. da~Silva, and Ana L.~C. Bazzan.
\newblock Using topological statistics to bias and accelerate route choice:
  preliminary findings in synthetic and real-world road networks.
\newblock In {\em Proceedings of Ninth International Workshop on Agents in
  Traffic and Transportation}, pages 1--8, New York, USA, 10 July 2016.

\bibitem{Oliveira+2017}
Thiago B.~F. Oliveira, Bruno C.~{\relax da} Silva, Stefanello F., Arthur
  Zachow, and Ana L.~C. Bazzan.
\newblock Extending a coupling metric for characterization of traffic networks:
  an application to the route choice problem.
\newblock In {\em Proc. of the 11th Workshop-School on Agents, Enviroments, and
  Applications (WESAAC 2017)}, S\~ao Paulo, May 2017.

\bibitem{Redwan&Bazzan2020}
Camil~S. Z. Redwan and Ana~L. C. Bazzan.
\newblock How hard is for agents to learn the user equilibrium? characterizing
  traffic networks by means of entropy.
\newblock {\em Advances in Complex Systems}, 23(04):2050011, 2020.

\bibitem{Tavares&Bazzan2012bwss}
Anderson~R. Tavares and Ana L.~C. Bazzan.
\newblock A multiagent based road pricing approach for urban traffic
  management.
\newblock In {\em Third Brazilian Workshop on Social Simulation}, pages
  99--105, 2012.

\bibitem{Huang+2020}
I-an Huang, Di~Mei, Anita Raja, Mohammad~Rashedul Hasan, and Ana~L.C. Bazzan.
\newblock Traffic optimization using a coordinated route updating mechanism.
\newblock {\em Int. J. of ITS}, 2020.
\newblock Submitted.

\bibitem{Santos&Bazzan2021}
Guilherme Dytz~{\relax dos} Santos and Ana L.~C. Bazzan.
\newblock Sharing diverse information gets driver agents to learn faster: an
  application in en route trip building.
\newblock {\em PeerJ Computer Science}, 7:e428, March 2021.

\bibitem{Bazzan&Kluegl2020}
A.~L.~C. Bazzan and F.~Kl\"{u}gl.
\newblock Experience sharing in a traffic scenario.
\newblock In Ivana Dusparic, Franziska Kl{\"{u}}gl, Marin Lujak, and Giuseppe
  Vizzari, editors, {\em Proc. of the 11th Int. Workshop on Agents in Traffic
  and Transportation}, volume 2701, pages 71--78, Santiago de Compostella,
  2020. CEUR-WS.org.

\bibitem{Bazzan+2008alamas}
Ana L.~C. Bazzan, Denise de~Oliveira, Franziska Kl\"ugl, and Kai Nagel.
\newblock Adapt or not to adapt -- consequences of adapting driver and traffic
  light agents.
\newblock In K.~Tuyls, A.~Now\'e, Z.~Guessoum, and D.~Kudenko, editors, {\em
  Adaptive Agents and Multi-Agent Systems III}, volume 4865 of {\em Lecture
  Notes in Artificial Intelligence}, pages 1--14. Springer-Verlag, 2008.

\bibitem{Lemos&Bazzan2019}
Liza~Lunardi Lemos and Ana L.~C. Bazzan.
\newblock Combining adaptation at supply and demand levels in microscopic
  simulation: a multiagent learning approach.
\newblock {\em Transportation Research Procedia}, 37:465--472, 2019.
\newblock Selected and peer-reviewed from the 21st EURO Working Group on
  Transportation (EWGT) meeting.

\bibitem{Lemos+2018}
Liza~Lunardi Lemos, Ana L.~C. Bazzan, and Marcia Pasin.
\newblock Co-adaptive reinforcement learning in microscopic traffic systems.
\newblock In {\em 2018 {IEEE} Congress on Evolutionary Computation, {CEC} 2018,
  Rio de Janeiro, Brazil, July 8-13, 2018}, pages 1--8, 2018.

\bibitem{Reymond&Nowe2019}
Mathieu Reymond and Ann Now\'e.
\newblock Pareto-dqn: Approximating the pareto front in complex multi-objective
  decision problems.
\newblock In {\em Proceedings of the Adaptive and Learning Agents Workshop 2019
  (ALA-19) at AAMAS}, 2019.

\bibitem{Auer+2002ucb}
Peter Auer, N.~Cesa-Bianchi, and Paul Fischer.
\newblock Finite-time analysis of the multiarmed bandit problem.
\newblock {\em Machine Learning}, 47(2/3):235--256, 2002.

\bibitem{Raith+2014}
Andrea Raith, Judith~YT Wang, Matthias Ehrgott, and Stuart~A Mitchell.
\newblock Solving multi-objective traffic assignment.
\newblock {\em Annals of Operations Research}, 222(1):483--516, 2014.

\end{thebibliography}

\end{document}